\documentclass[10pt]{article}
\usepackage{feynmf}
\usepackage{epsfig}

\begin{document}

\title{Exchange Current Operators and Electromagnetic Dipole Transitions in 
Heavy Quarkonia}

\author{T.A. L\"ahde$\,$\footnote{e-mail: talahde@pcu.helsinki.fi}
\\ \vspace{0.3cm} \\
{\normalsize \it Helsinki Institute of Physics and Department of 
Physical Sciences,}\\
{\normalsize \it PL 64 University of Helsinki, 00014 Finland} }
\date{}
\maketitle
\thispagestyle{empty}

\begin{abstract}
The electromagnetic E1 and M1 transitions in heavy quarkonia~($c\bar 
c$\,,\,$b\bar b$\,,\,$c\bar b$) and the magnetic moment of the $B_c^\pm$ 
are calculated within the framework of the covariant Blankenbecler-Sugar 
(BSLT) equation. The aim of this paper is to study the effects of two-quark 
exchange current operators which involve the $Q\bar Q$ interaction, that 
arise in the BSLT (or Schr\"odinger) reduction of the Bethe-Salpeter 
equation. These are found to be small for E1 dominated transitions such as 
$\psi(nS)\rightarrow \chi_{cJ}\,\gamma$ and $\Upsilon(nS)\rightarrow 
\chi_{bJ}\,\gamma$, but significant for the M1 dominated ones. It 
is shown that a satisfactory description of the empirical data on E1 and M1 
transitions in charmonium and bottomonium requires unapproximated 
treatment of the Dirac currents of the quarks. Finally, it is demonstrated 
that many of the transitions are sensitive to the form of the $Q\bar Q$ 
wavefunctions, and thus require a realistic treatment of the large 
hyperfine splittings in the heavy quarkonium systems.
\end{abstract}
\newpage

\section{Introduction}

The radiative decays of heavy quarkonia ($c\bar c$, $b\bar b$, 
$c\bar b$) have drawn much interest, as they can provide direct information 
on both the heavy quarkonium wavefunctions and the $Q\bar Q$ interaction. 
Theoretical descriptions range from potential 
models~\cite{McClary}~-~\cite{Quigg}, QCD sum 
rules~\cite{Colangelo,Kiselev}, Heavy Quark Effective Theory 
(HQET)~\cite{Casa} to Non-Relativistic QCD (NRQCD)~\cite{Bramb}. As 
experimental data of a reasonable quality exists for a number of 
transitions in the $c\bar c$ and $b\bar b$ systems~\cite{PDG,Hagiwara}, a 
fair assessment of the quality of model predictions is already possible. 
The measured $\gamma$ decays in the charmonium~($c\bar c$) system include 
the E1 transitions $\chi_{cJ} \rightarrow J/\psi\,\gamma$ and 
$\psi\,'\rightarrow \chi_{cJ}\,\gamma$, as 
well as the spin-flip M1 transitions $J/\psi \rightarrow \eta_c\gamma$ and 
$\psi\,'\rightarrow \eta_c\gamma$. The situation in the bottomonium~($b\bar 
b$) system is, however, less satisfactory as the total widths of the 
$\chi_{bJ}$ states are not known, and none of the spin-flip M1 decays 
observed.

Previous calculations of the widths for E1 transitions in heavy 
quarkonia~\cite{McClary,GrOld,Snellman} have acheived qualitative agreement 
with experiment. However, the situation concerning the M1 transitions has 
remained unsatisfactory for a long time~\cite{Shifmanpap} as the width for 
$J/\psi \rightarrow \eta_c\gamma$ has typically been overpredicted by a 
factor of~$\sim 3$. Calculations of M1 widths using the nonrelativistic 
Schr\"odinger equation~\cite{GrOld,Oldlahde} have demonstrated that the M1 
transitions in charmonium are sensitive both to the relativistic aspects of 
the spin-flip operator as well as the Lorentz structure of the $Q\bar Q$ 
interaction. The results obtained in ref.~\cite{Oldlahde} suggest that a 
scalar confining interaction may explain the observed width of $\sim 1$~keV 
for $J/\psi\rightarrow \eta_c\gamma$, provided that an unapproximated 
expression for the single quark spin-flip operator is used. However, as 
many of the M1 transitions are very sensitive to the exact form of the 
spin-flip operator and the $Q\bar Q$ wavefunctions, a more realistic model 
for the wavefunctions of heavy quarkonia is called for.

This paper reports a comprehensive calculation of the E1 and M1 
transitions in heavy quarkonia, as well as the magnetic moment of the 
$B_c^\pm$~($c\bar b,\,b\bar c$) meson, within the framework of the 
covariant Blankenbecler-Sugar (BSLT) equation~\cite{Blank}. The $Q\bar Q$ 
interaction Hamiltonian is formed of scalar confining and vector one-gluon 
exchange (OGE) components, in addition to a small instanton induced 
interaction~\cite{Zahed}. The hyperfine components of the $Q\bar Q$ 
interaction have been fully taken into account, which is found to be 
important for several E1 and M1 transitions, in line with the conclusion 
reached in ref.~\cite{GrE1bb}. On the other hand, the elimination of the 
negative energy components of the Bethe-Salpeter equation leads to the 
appearance of two-quark transition operators~\cite{Coester}. These exchange 
current operators, which depend explicitly on the Lorentz coupling 
structure of the $Q\bar Q$ interaction, have been shown to give large 
corrections to the single quark transition 
operators~\cite{Oldlahde,LahdeDs}.

The aim of this paper is to evaluate the exchange current contributions to 
the electric and magnetic dipole operators for the $Q\bar Q$ interaction 
Hamiltonian described above. The contributions from exchange charge 
operators to the E1 transition rates are shown to be highly suppressed by 
the large masses of the charm and bottom quarks. However, it is found that 
the exchange magnetic moment operator associated with the scalar confining 
component of the $Q\bar Q$ interaction gives a large contribution, which 
makes it possible to bring the calculated width for $J/\psi\rightarrow 
\eta_c\gamma$ into agreement with the observed width of $\sim 1$~keV. It is 
also found that the empirical width of $\sim 1$ keV for the forbidden 
transition $\psi\,'\rightarrow \eta_c\gamma$ may be similarly explained if a 
relativistic single-quark spin-flip operator is employed togehter with 
wavefunctions that model the spin-spin splitting of the $\psi$ states. This 
provides evidence in favor of a dominant scalar Lorentz structure for the 
effective linear confining interaction, as an effective vector interaction 
gives a vanishing contribution to the spin-flip operator for M1 transitions 
in the $c\bar c$ and $b\bar b$ systems~\cite{Oldlahde}. On the other hand, 
for the $B_c^\pm$ the OGE interaction contributes an exchange magnetic 
moment operator that counteracts that from the scalar confining 
interaction. 

The layout of this paper is as follows: Section 2 presents the transition 
operators for E1 and M1 decay, while section 3 deals with the Hamiltonian 
model and $Q\bar Q$ wavefunctions, along with formulas for the E1 and M1 
widths. Section 4 presents the numerical results for the radiative decays 
and the $B_c^\pm$ magnetic moment, and section 5 contains a discussion of 
the obtained results.

\newpage

\section{The Electric Dipole and Magnetic Moment Operators}

\subsection{The Charge Density and Electric Dipole Operators}
\label{E1op}

From the \mbox{$S$-matrix} element for one-photon emission by a
two-quark system,
\begin{equation}
S_{fi}\:=\:-i\:\left<f\right|\:\delta_1\delta_2\: 
\hat\varepsilon_\mu\:J_\mu \left|i\right>, \label{ampl}
\end{equation}
where $\delta_{1,2}$ are four-momentum conserving delta functions for 
quarks 1 and 2, and $J_\mu$ is the current operator of the two-quark 
system, the electromagnetic transition amplitude for $Q\bar Q$ systems is 
obtained, in the impulse approximation, as
\begin{equation}
T_{fi}\:=\:-\int d^3r_1 d^3r_2\: 
\varphi_f^*(\vec r_1,\vec r_2)\:
\hat\varepsilon\cdot\left[e^{i\vec q\cdot\vec r_1}\,\vec 
\jmath_1(\vec q\,) + e^{i\vec q\cdot\vec r_2}\,\vec \jmath_2(\vec 
q\,)\right] \varphi_i(\vec r_1,\vec r_2), \label{matr}
\end{equation}
where $\vec q$ and $\hat\varepsilon$ denote the 
momentum and polarization of the emitted photon, respectively, while 
$\varphi_i$ and $\varphi_f$ denote the orbital wavefunctions of the 
initial and final heavy quarkonium states. In the above equation, 
$\vec\jmath_1$ and $\vec\jmath_2$ denote the single quark current operators 
of quarks 1 and 2, respectively. Note that the current operator 
$\vec\jmath\,(\vec q\,)$ corresponds to the quantity in square brackets in 
eq.~(\ref{matr}). The amplitude for a $\gamma$ transition is then
\begin{equation}
T_{fi}\:=\: i\,|\vec q\,|\,\int d^3r_1 d^3r_2 \: 
\varphi_f^*(\vec r_1,\vec r_2)\: 
\hat\varepsilon\cdot\vec d\,(\vec r_1,\vec r_2)
\,\varphi_i(\vec r_1,\vec r_2),\label{matr2}
\end{equation}
where the dipole operator $\vec d(\vec r_1,\vec r_2)$ is of the form
\begin{equation}
\vec d\,(\vec r_1,\vec r_2) = \int d^3r' e^{i\vec q\cdot\vec r\,'}
\vec r\,'\, \rho(\vec r\,',\vec r_1,\vec r_2).
\end{equation}
In order to distinguish the above model from the rigorous E1 
approximation, it will be referred to as "dynamical" throughout this paper. 
The E1 approximation is then obtained in the limit $\vec q \rightarrow 0$. 
If contributions to the charge operator $\rho(\vec r\,',\vec r\,)$, that 
are proportional to higher powers of the photon momentum are to be 
included, then the usefulness of the dynamical model is apparent. More 
significantly, the dynamical model allows the recoil of the heavy meson to 
be accounted for. 

\begin{figure}[h!]
\begin{center}
\begin{tabular}{c c}
\begin{fmffile}{ex1}
\begin{fmfgraph*}(160,130) \fmfpen{thin}
\fmfcmd{%
 vardef port (expr t, p) =
  (direction t of p rotated 90)
   / abs (direction t of p)
 enddef;}  
\fmfcmd{%
 vardef portpath (expr a, b, p) =
  save l; numeric l; l = length p;
  for t=0 step 0.1 until l+0.05:
   if t>0: .. fi point t of p
    shifted ((a+b*sind(180t/l))*port(t,p))
  endfor
  if cycle p: .. cycle fi
 enddef;}
\fmfcmd{%
 style_def brown_muck expr p =
  shadedraw(portpath(thick/2,2thick,p)
   ..reverse(portpath(-thick/2,-2thick,p))
   ..cycle)
 enddef;}
\fmfleft{i2,i1}
\fmfright{o2,o1}
\fmftop{o3}
\fmf{fermion,label=$p_1$}{i1,v3}
\fmf{fermion,label=$p_a$}{v3,v1}
\fmf{fermion,label=$p_1'$}{v1,o1}
\fmf{fermion,label=$p_2'$}{o2,v2}
\fmf{fermion,label=$p_2$}{v2,i2}
\fmf{photon,label=$q$,label.side=right}{v3,o3}
\fmf{brown_muck,lab.s=right,lab.d=4thick,lab=$V_{Q\bar Q}(k_2)$,label.side=right}{v1,v2}
\fmfdot{v1,v2,v3}
\fmfforce{(.1w,.85h)}{i1}
\fmfforce{(.9w,.85h)}{o1}
\fmfforce{(.1w,.15h)}{i2}
\fmfforce{(.9w,.15h)}{o2}
\fmfforce{(.2w,.h)}{o3}
\fmfforce{(.5w,.70h)}{v1}
\fmfforce{(.5w,.30h)}{v2}
\fmfforce{(.3w,.80h)}{v3}
\end{fmfgraph*}
\end{fmffile}
&
\begin{fmffile}{ex2}
\begin{fmfgraph*}(160,130) \fmfpen{thin}
\fmfcmd{%
 vardef port (expr t, p) =
  (direction t of p rotated 90)
   / abs (direction t of p)
 enddef;}  
\fmfcmd{%
 vardef portpath (expr a, b, p) =
  save l; numeric l; l = length p;
  for t=0 step 0.1 until l+0.05:
   if t>0: .. fi point t of p
    shifted ((a+b*sind(180t/l))*port(t,p))
  endfor
  if cycle p: .. cycle fi
 enddef;}
\fmfcmd{%
 style_def brown_muck expr p =
  shadedraw(portpath(thick/2,2thick,p)
   ..reverse(portpath(-thick/2,-2thick,p))
   ..cycle)
 enddef;}
\fmfleft{i2,i1}
\fmfright{o2,o1}
\fmftop{o3}
\fmf{fermion,label=$p_1$}{i1,v1}
\fmf{fermion,label=$p_b$}{v1,v3}
\fmf{fermion,label=$p_1'$}{v3,o1}
\fmf{fermion,label=$p_2'$}{o2,v2}
\fmf{fermion,label=$p_2$}{v2,i2}
\fmf{photon,label=$q$,label.side=right}{v3,o3}
\fmf{brown_muck,lab.s=right,lab.d=4thick,lab=$V_{Q\bar Q}(k_2)$,label.side=right}{v1,v2}
\fmfdot{v1,v2,v3}
\fmfforce{(.1w,.85h)}{i1}
\fmfforce{(.9w,.85h)}{o1}
\fmfforce{(.1w,.15h)}{i2}
\fmfforce{(.9w,.15h)}{o2}
\fmfforce{(.6w,.h)}{o3}
\fmfforce{(.5w,.70h)}{v1}
\fmfforce{(.5w,.30h)}{v2}
\fmfforce{(.7w,.80h)}{v3}
\end{fmfgraph*}
\end{fmffile}
\\ & \\
\begin{fmffile}{ex3}
\begin{fmfgraph*}(160,130) \fmfpen{thin}
\fmfcmd{%
 vardef port (expr t, p) =   
  (direction t of p rotated 90)
   / abs (direction t of p)
 enddef;}
\fmfcmd{%
 vardef portpath (expr a, b, p) =
  save l; numeric l; l = length p;
  for t=0 step 0.1 until l+0.05:
   if t>0: .. fi point t of p
    shifted ((a+b*sind(180t/l))*port(t,p))
  endfor
  if cycle p: .. cycle fi
 enddef;}
\fmfcmd{%
 style_def brown_muck expr p =
  shadedraw(portpath(thick/2,2thick,p)  
   ..reverse(portpath(-thick/2,-2thick,p))
   ..cycle)
 enddef;}
\fmfleft{i2,i1}
\fmfright{o2,o1}  
\fmftop{o3}   
\fmf{fermion,label=$p_1$}{i1,v3}
\fmf{fermion,label=$p_a$}{v3,v1}
\fmf{fermion,label=$p_1'$}{v1,o1}
\fmf{fermion,label=$p_2'$}{o2,v2}
\fmf{fermion,label=$p_2$}{v2,i2}
\fmf{photon,label=$q$,label.side=right}{v3,o3}  
\fmf{brown_muck,lab.s=right,lab.d=4thick,lab=$V_{Q\bar Q}(k_2)$,label.side=right}{v1,v2}
\fmfdot{v1,v2,v3}
\fmfforce{(.1w,.85h)}{i1}
\fmfforce{(.9w,.75h)}{o1}
\fmfforce{(.1w,.05h)}{i2}
\fmfforce{(.9w,.05h)}{o2}
\fmfforce{(.5w,.60h)}{v1}
\fmfforce{(.5w,.20h)}{v2}
\fmfforce{(.6w,.80h)}{v3}
\fmfforce{(.5w,.h)}{o3}  
\end{fmfgraph*}
\end{fmffile}
&
\begin{fmffile}{ex4}
\begin{fmfgraph*}(160,130) \fmfpen{thin}  
\fmfcmd{%
 vardef port (expr t, p) =
  (direction t of p rotated 90)
   / abs (direction t of p)
 enddef;}     
\fmfcmd{%
 vardef portpath (expr a, b, p) =
  save l; numeric l; l = length p;
  for t=0 step 0.1 until l+0.05: 
   if t>0: .. fi point t of p   
    shifted ((a+b*sind(180t/l))*port(t,p))
  endfor
  if cycle p: .. cycle fi
 enddef;}
\fmfcmd{%
 style_def brown_muck expr p =
  shadedraw(portpath(thick/2,2thick,p)
   ..reverse(portpath(-thick/2,-2thick,p))
   ..cycle)
 enddef;}
\fmfleft{i2,i1}
\fmfright{o2,o1}
\fmftop{o3}  
\fmf{fermion,label=$p_1$}{i1,v1}
\fmf{fermion,label=$p_b$,label.side=right}{v1,v3}
\fmf{fermion,label=$p_1'$}{v3,o1}
\fmf{fermion,label=$p_2'$}{o2,v2}
\fmf{fermion,label=$p_2$}{v2,i2}
\fmf{photon,label=$q$,label.side=right}{v3,o3}  
\fmf{brown_muck,lab.s=right,lab.d=4thick,lab=$V_{Q\bar Q}(k_2)$,label.side=right}{v1,v2}
\fmfdot{v1,v2,v3}
\fmfforce{(.1w,.75h)}{i1}
\fmfforce{(.9w,.85h)}{o1}
\fmfforce{(.1w,.05h)}{i2}
\fmfforce{(.9w,.05h)}{o2}
\fmfforce{(.5w,.60h)}{v1}
\fmfforce{(.5w,.20h)}{v2}
\fmfforce{(.4w,.80h)}{v3}
\fmfforce{(.3w,.h)}{o3}  
\end{fmfgraph*}
\end{fmffile} \\
\end{tabular}
\caption{Relativistic Born diagrams for photon emission by a heavy 
constituent quark. The lower diagrams describe the negative energy 
components of the upper diagrams, and can be obtained from the latter by 
separation of the intermediate quark propagators $p_a$ and $p_b$ into 
negative and positive energy components. Note that similar diagrams 
describe photon emission by the heavy antiquark. The exchange charge 
operators that correspond to the above diagrams have been calculated for 
different Lorentz coupling structures of the interaction $V$ in 
ref.~\cite{Helminen}, which in the case of the $Q\bar Q$ 
interaction will contain scalar confining and vector OGE components.} 
\label{feyn}
\end{center}
\end{figure}
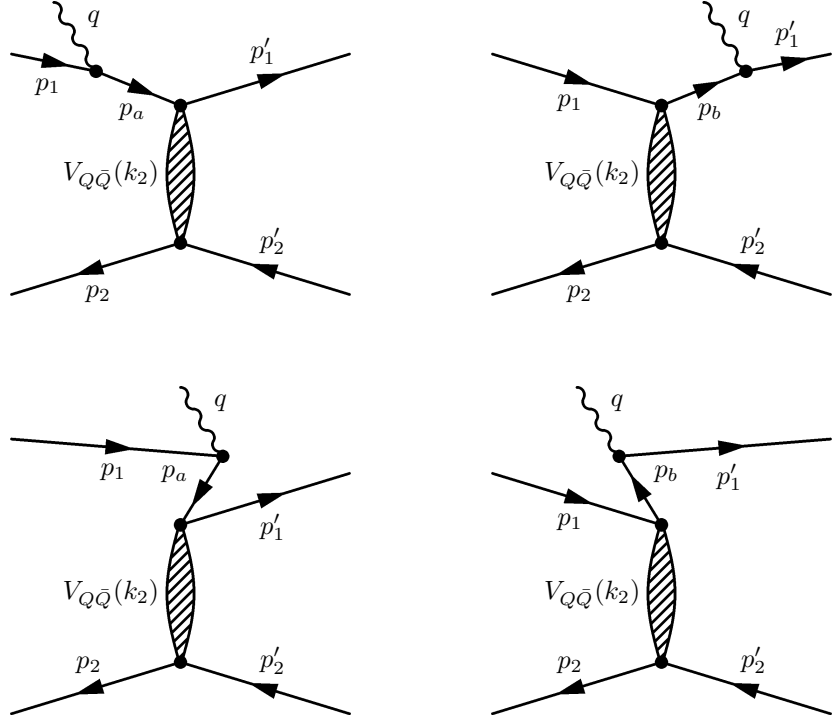

The charge density operator $\rho(\vec r\,')$ contains, in addition to the 
single quark contribution $\rho_{\mathrm{sq}}$, an exchange part 
$\rho_{\mathrm{ex}}$, which arises from processes that are illustrated by 
the diagrams shown in Fig.~\ref{feyn}. These contributions arise from the 
elimination of the negative energy components in the reduction of the 
Bethe-Salpeter equation to a BSLT (or Schr\"odinger) equation. An obvious 
constraint is that two-quark contributions to the charge density should 
have vanishing volume integrals.

Consider first the single quark charge operator $\rho_{\mathrm{sq}}(\vec 
r\,',\vec r\,) = \rho_1(\vec r\,',\vec r_1) + \rho_2(\vec r\,',\vec r_2)$.
The corresponding dipole operator may be expressed as
\begin{equation}
\vec d\:_{\mathrm{sq}}(\vec r_1,\vec r_2) = Q_1\,\vec r_1\:e^{i\vec 
q_f\cdot\vec r_1\,} +  Q_2\,\vec r_2\:e^{i\vec q_f\cdot\vec r_2\,}, 
\label{dip}
\end{equation}
where $Q_1$ and $Q_2$ are the appropriate electric charges of the 
constituent charm and bottom quarks. Note that $Q_1$ is taken to be the 
charge of the heavy quark, while $Q_2$ denotes that of the heavy antiquark.
Relativistic modifications to the above expression will be considered 
further on. Insertion of eq.~(\ref{dip}) into eq.~(\ref{matr2}) yields the 
amplitude
\begin{equation}
T_{fi}\:=\: i\,|\vec q_f|\:(2\pi)^3\,\delta^3(P_f - P_i - q_f)
\int d^3r \:\varphi_f^*(\vec r\,)\:
\hat\varepsilon\cdot\vec d\,(\vec r\,)\,\varphi_i(\vec r\,),
\label{matr2.5}
\end{equation}
where $\vec r = \vec r_1 - \vec r_2$, to which the contribution from the 
single quark dipole operator~(\ref{dip}) is
\begin{equation}
\vec d\:_{\mathrm{sq}}(\vec r\,) = \left[\frac{Q_1m_2 - 
Q_2m_1}{m_1+m_2}\right]\:\vec r\:e^{i\vec q_f\cdot\vec r/2\,},
\label{cmdip}
\end{equation}
where $\vec r = \vec r_1 - \vec r_2$. Again, the E1 expression~\cite{Quigg} 
is obtained by dropping the exponential in eq.~(\ref{cmdip}) and setting 
$\vec q_f = 0$ in the delta function in eq.~(\ref{matr2.5}). Note that for 
the $c\bar c$ and $b\bar b$ systems, the factor in brackets in 
eq.~(\ref{cmdip}) reduces to the charge of the charm and bottom quark, 
respectively.

Consider next the two-quark exchange charge operators from the Born 
diagrams given in Fig.~\ref{feyn}. If those operators are decomposed 
according to $\rho_{\mathrm{ex}}(\vec r\,',\vec r_1,\vec r_2) =
\rho_{\mathrm{ex}1}(\vec r\,',\vec r_1) + \rho_{\mathrm{ex}2}(\vec 
r\,',\vec r_2)$, then the exchange charge contribution to the two-quark 
dipole operator 
\begin{equation}
\vec d_{\mathrm{ex}}\,(\vec r_1,\vec r_2) = \int d^3r' e^{i\vec
q_f\cdot\vec r\,'}\vec r\,'\,\rho_{\mathrm{ex}}(\vec r\,',\vec r_1,\vec r_2)
\end{equation}
may be expressed as
\begin{equation}
\vec d_{\mathrm{ex1}}\,(\vec r_1) = \vec r_1\,e^{i\vec 
q_f\cdot\vec r_1}\!\!\int\frac{d^3k_2}{(2\pi)^3}\,
e^{-i\vec k_2\cdot\vec r}\,\rho_{\mathrm{ex}1}(\vec q_f,\vec k_2)
-\lim_{q\rightarrow q_f}\left[e^{i\vec q\cdot\vec r_1}\,i\nabla_{\vec q}
\!\int\frac{d^3k_2}{(2\pi)^3}\,e^{-i\vec k_2\cdot\vec r}\,
\rho_{\mathrm{ex}1}(\vec q,\vec k_2)\right], \label{exop}
\end{equation}
from which the E1 approximation can again be obtained by setting $q_f 
\rightarrow 0$. 

Having established eq.~(\ref{exop}), the relativistic modifications to the 
single quark charge operator as well as the two-quark exchange charge 
operators from Fig.~\ref{feyn} may be considered. To second order in $v/c$, 
the single quark charge operator may be expressed in the 
form~\cite{Helminen}
\begin{equation}
\rho_{\mathrm{sq}} \simeq Q_1\left[1 - \frac{\vec q\,^2}{8m^2} + 
\frac{i\vec\sigma_1\cdot\vec p\,'_1\times\vec p_1}{4m^2}\:\right] + 
(1\rightarrow 2), \label{rsq}
\end{equation}
where it is understood that the contribution from quark 2 is obtained by 
replacing the indices accordingly. The second term on the 
r.h.s. is the relativistic Darwin-Foldy term. It will be shown that the 
effect of this term is very small because of the large masses of the heavy 
constituent quarks. Note that this expansion is justified by the small 
coefficient of the $q^2$ term; It has been shown in ref.~\cite{Oldlahde} 
that such an expansion cannot be used for the magnetic moment operator. The 
spin-orbit term in eq.~(\ref{rsq}) is linear in the photon momentum $\vec 
q$ and will be left out in this work because of its smallness. For 
transitions between $S$-wave states, this term vanishes 
entirely~\cite{Helminen}. Note that in the E1 approximation, the 
contribution from the Darwin-Foldy term to the dipole operator will 
likewise vanish.

The exchange charge density operators that are associated with the $Q\bar 
Q$ interaction have been calculated in \mbox{ref.~\cite{Helminen},} where 
the appropriate operators were extracted for different Lorentz invariants 
for systems composed of quarks with equal masses. When generalized to the 
case of unequal quark masses, the required operators are obtained as
\begin{equation}
\rho_{\mathrm{ex}}^{\,\mathrm c} = \frac{Q_1}{4m_1^{\,3}}\,q^2\,V_c(\vec 
k_2\,) + \frac{Q_2}{4m_2^{\,3}}\,q^2\,V_c(\vec k_1\,)
\label{rconf}
\end{equation}
for the scalar confining interaction, and
\begin{equation}
\rho_{\mathrm{ex}}^{\,\mathrm g} = 
\frac{Q_1}{4m_1^{\,2}}
\left[\frac{\vec q\cdot\vec k_2}{m_1} + \frac{2}{3}\,
\frac{\vec q\cdot\vec k_2\:\vec\sigma_1\cdot\vec\sigma_2}{m_2}\right] 
V_g(\vec k_2\,) +
\frac{Q_2}{4m_2^{\,2}}
\left[\frac{\vec q\cdot\vec k_1}{m_2} + \frac{2}{3}\,
\frac{\vec q\cdot\vec k_1\:\vec\sigma_1\cdot\vec\sigma_2}{m_1}\right] 
V_g(\vec k_1\,) 
\label{roge}
\end{equation}
for the vector coupled OGE interaction. In the above expressions, $V_c$ and 
$V_g$ denote the momentum space forms of the confining and OGE 
interactions, respectively. Note that the terms in eqs.~(\ref{rconf}) 
and~(\ref{roge}) that depend on $\vec k_2$ correspond to the contribution 
$\rho_{\mathrm{ex}1}$ in eq.~(\ref{exop}) and vice versa. In the E1 
approximation, the contribution to the dipole operator from 
eq.~(\ref{rconf}) vanishes, while in the dynamical model, evaluation of 
eq.~(\ref{exop}) yields, analogously to eq.~(\ref{cmdip}),
\begin{equation}
\vec d\:_{\mathrm{ex}}^{\mathrm{Conf}}(\vec r\,) = q_f^2\,
\left[\frac{Q_1}{4m_1^{\,3}}\,\frac{m_2}{m_1+m_2} - 
\frac{Q_2}{4m_2^{\,3}}\,\frac{m_1}{m_1+m_2}\right]\:
\vec r\:\:V_c(r)\:\:e^{i\vec q_f\cdot\vec r/2\,}, \label{cdip}
\end{equation}
where the scalar confining interaction is of the form $V_c(r) = cr$. Note 
that the second term in eq.~(\ref{exop}) vanishes because $\vec 
q\cdot\hat\varepsilon = 0$. On the other hand, the OGE 
expression~(\ref{roge}) gives a contribution, the dominant term of which 
is 
\begin{equation}
\vec d\:_{\mathrm{ex}}^{\mathrm{Oge}}(\vec r\,) = 
\left[\frac{Q_1}{4m_1^{\,2}}\left(\frac{1}{m_1} + 
\frac{2}{3}\frac{\vec\sigma_1\!\cdot\vec\sigma_2}{m_2}\right)
-\frac{Q_2}{4m_2^{\,2}}\left(\frac{1}{m_2} +
\frac{2}{3}\frac{\vec\sigma_1\!\cdot\vec\sigma_2}{m_1}\right)
\right]\:
\vec r\:\left(\frac{\partial V_g(r)}{r\,\partial r}\right)\,
e^{i\vec q_f\cdot\vec r/2\,}. \label{gdip}
\end{equation}
Here $V_g(r)$ denotes the form of the OGE interaction in configuration 
space, and is here taken to be the Fourier transform of eq.~(\ref{OGEpot}). 
This choice allows the inclusion of the running coupling of QCD. Note that 
eq.~(\ref{gdip}) gives a non-vanishing contribution also in the E1 
approximation.

\pagebreak
\subsection{The Current Density and Magnetic Moment Operators}

In the impulse approximation, the spin-flip magnetic moment operator for M1 
transitions between $S$-wave heavy quarkonium states may be obtained from 
the amplitude
\begin{equation}
T_{fi}\:=\:-(2\pi)^3\,\delta^3(P_f - P_i - q_f)
\int d^3r\: \varphi_f^*(\vec r\,)
\:\hat\varepsilon\cdot\left[e^{i\vec q\cdot\vec r/2}\,\vec 
\jmath_1(\vec q\,) + e^{-i\vec q\cdot\vec r/2}\,\vec 
\jmath_2(\vec q\,)\right] \varphi_i(\vec r\,), 
\label{matrM1}
\end{equation}
where $\vec r = \vec r_1 - \vec r_2$. The corresponding matrix element for 
an M1 transition may be written in the form
\begin{equation}
{\cal M}_{fi} =\,i\int d^3r \: \varphi_f^*(\vec r\,)\:\:\vec q\times
\hat\varepsilon\cdot\vec \mu_{\mathrm{sf}}\:\:\varphi_i(\vec r\,),
\label{matr3}
\end{equation}
where $\vec\mu_{\mathrm{sf}}$ denotes the spin-flip part of the 
standard magnetic moment operator. In addition to the single quark current 
operators in eq.~(\ref{matrM1}), the negative energy Born diagrams in 
Fig.~\ref{feyn} contribute a two-quark current operator 
$\vec\jmath_{\mathrm{ex}}$. Decomposition of the single quark current into 
contributions from quark~1 and~2 according to 
$\vec\jmath_{\mathrm{sq}} = \vec\jmath_1 + \vec\jmath_2$ yields the 
magnetic moment operator
\begin{equation}
\vec \mu_{\mathrm{sq}}\:=\:\lim_{\vec q \rightarrow 
0}\left[-\frac{i}{2}\:\nabla_q\times
\left(e^{i\vec q\cdot\vec r/2}\,\vec\jmath_1(\vec q\,) + 
e^{-i\vec q\cdot\vec r/2}\,\vec\jmath_2(\vec q\,)\right)\:\right],
\label{musq}
\end{equation}
in the nonrelativistic impulse approximation (NRIA). However, previous work 
has demonstrated that the static magnetic moment operators of the baryons 
are significantly modified by the canonical boosts of the constituent 
quark spinors~\cite{Dannbom,Helm-PhD}. In ref.~\cite{Oldlahde}, it was 
shown that the static spin-flip magnetic moment operators for M1 decay of 
$Q\bar Q$ states are also significantly affected, despite the large masses 
of the charm and bottom constituent quarks. 

The matrix element that corresponds to eq.~(\ref{matr3}) in the 
relativistic impulse approximation (RIA) may be obtained as
\begin{equation}
{\cal M}_{fi}^{\mathrm{Rel}} =\,i\int\frac{d^3P}{(2\pi)^3}\,d^3r\,d^3r' 
\:e^{i\vec P\cdot (\vec r\,' - \vec r\,)}\: 
\varphi_f^*(\vec r\,')\:\:\vec q\times
\hat\varepsilon\cdot\vec \mu\,_{\mathrm{sq}}^{\mathrm{Rel}}(\vec P\,)
\:\:\varphi_i(\vec r\,),\label{matr4}
\end{equation}
where the final and initial state coordinates $\vec r\,'$ and $\vec r$ are 
defined as $\vec r\,'_1 - \vec r\,'_2$ and $\vec r_1 - \vec r_2$ 
respectively. In eq.~(\ref{matr4}), the momentum variable $\vec P$ is 
defined as $\vec P = (\vec p\,' + \vec p\,)/2$, where $\vec p\,'$ and 
$\vec p$ are the relative momenta in the representation $\vec p_1 = \vec 
P_i/2 + \vec p$, $\vec p_2 = \vec P_i/2 - \vec p\:$ and $\:\vec p\,'_1 = 
\vec P_f/2 + \vec p\,'$, $\vec p\,'_2 = \vec P_f/2 - \vec p\,'$. 
The relativistic single quark magnetic moment operator that appears in the 
matrix element~(\ref{matr4}) can be obtained from eq.~(\ref{musq}) by the 
substitution $\vec r \rightarrow (\vec r\,' + \vec r\,)/2$.

In the nonrelativistic case, the spin-dependent part of the single quark 
current operator $\vec\jmath_{\mathrm{sq}} = \vec\jmath_1 + \vec\jmath_2$ 
is of the form
\begin{equation}
\vec\jmath\:\:_{\mathrm{sq}}^{\mathrm{spin}} = \frac{ie}{2}\,(\vec\sigma_1 
+ \vec\sigma_2)\times\vec q\,\left[\frac{Q_1}{2m_1} + \frac{Q_2}{2m_2}\right] 
+ \frac{ie}{2}\,(\vec\sigma_1 - \vec\sigma_2)
\times\vec q\,\left[\frac{Q_1}{2m_1} - \frac{Q_2}{2m_2}\right].
\label{nrscurr}
\end{equation}
The first term, which vanishes for equal mass quarkonia, describes the 
magnetic moment of the two-quark system whereas the second term is the 
spin-flip operator for M1 decay in the nonrelativistic impulse 
approximation (NRIA). The corresponding spin-flip operator in the 
relativistic impulse approximation (RIA) has been calculated in 
refs.~\cite{Helm-PhD,LahdeDs}, and may for transitions between $S$-wave 
states be expressed as
\begin{equation}
\vec \mu\,_{\mathrm{sq}}^{\mathrm{Rel}}\:=\:\frac{e}{2}\left[
\frac{Q_1}{2m_1}f_1^\gamma - 
\frac{Q_2}{2m_2}f_2^\gamma\right]\:(\vec\sigma_1 - \vec\sigma_2), 
\hspace{1.5cm}
f_i^\gamma = \frac{m_i}{3E_i}\left[2+\frac{m_i}{E_i}\right],
\label{RIA}
\end{equation}
with $E_i = \sqrt{P^2 + m_i^2}$. It is apparent from the above expression 
that the relativistic treatment will effectively weaken the NRIA result.

The relativistic Born diagrams in Fig.~\ref{feyn} contribute significant 
two-quark exchange currents that give rise to two-quark magnetic moment 
operators~\cite{Tsushima}. This situation is akin to that for the magnetic 
moments of the \mbox{baryons~\cite{Dannbom,Helm-PhD},} but in that case 
additional complications arise from flavor dependent meson exchange 
interactions. Decomposition of the exchange current operator according to
$\vec\jmath_{\mathrm{ex}}(\vec 
q,\vec k_1,\vec k_2) = \vec\jmath_{\mathrm{ex1}}(\vec q,\vec k_2)
+\vec\jmath_{\mathrm{ex2}}(\vec q,\vec k_1)$ yields the exchange magnetic 
moment operator
\begin{equation}
\vec \mu_{\mathrm{ex}}\:=\:\lim_{\vec q \rightarrow 
0}\left[-\frac{i}{2}\:\nabla_q\times
\left(e^{i\vec q\cdot\vec r/2}\!\int \frac{d^3k_2}{(2\pi)^3}\,
e^{-i\vec k_2\cdot\vec r}\,\vec\jmath_{\mathrm{ex1}}(\vec q,\vec k_2) +
e^{-i\vec q\cdot\vec r/2}\!\int \frac{d^3k_1}{(2\pi)^3}\,
e^{i\vec k_1\cdot\vec r}\,\vec\jmath_{\mathrm{ex2}}(\vec q,\vec k_1)
\right)\:\right].   
\label{muex}
\end{equation}
However, the exchange magnetic moment operators turn out to be difficult to 
calculate directly from the above equation. Instead, they can be 
conveniently extracted from the expression~\cite{Tsushima}
\begin{equation}
\vec \mu_{\mathrm{ex}}\:=\:\lim_{\vec q \rightarrow 0}
\left[\,-\frac{i}{2}\,\int\!\frac{d^3k}{(2\pi)^3}\,e^{-i\vec k\cdot\vec 
r}\:\nabla_q\times\left\{
\vec\jmath_{\mathrm{ex1}}\left(\frac{\vec q}{2} + \vec k\,\right) + 
\vec\jmath_{\mathrm{ex2}}\left(\frac{\vec q}{2} - \vec k\,\right)
\right\}\right], \label{muex2}
\end{equation}
where $\vec k_2 = \vec k + \vec q/2$ and $\vec k_1 = -\vec k + \vec q/2$. 
By means of eq.~(\ref{muex2}), it is now possible to consider the two-quark 
current operators for the scalar confining and vector OGE interactions, as 
calculated from the diagrams in Fig.~\ref{feyn} by 
refs.~\cite{LahdeDs,Tsushima}. The two-quark current operator associated 
with the scalar confining interaction is of the form
\begin{equation}
\vec\jmath_{\mathrm{ex}}^{\:\:\mathrm{c}} \:\:=\:\: 
-e\,\left(\frac{Q_1^*\vec P_1}{m_1^2} + \frac{Q_2^*\vec P_2}{m_2^2}
+ \frac{i}{2}(\vec\sigma_1+\vec\sigma_2)\times\vec q\left[
\frac{Q_1^*}{2m_1^2} + \frac{Q_2^*}{2m_2^2}\right] +
\frac{i}{2}(\vec\sigma_1-\vec\sigma_2)\times\vec q\left[
\frac{Q_1^*}{2m_1^2} - \frac{Q_2^*}{2m_2^2}\right]\right), \label{2qcurr}
\end{equation}  
in which case the two-quark magnetic moment operator is most 
conveniently computed using eq.~(\ref{muex}), as the spin part of 
eq.~(\ref{2qcurr}) depends explicitly on the photon momentum $\vec q$. In 
the above equation, the variables $Q_1^*$ and $Q_2^*$ are defined as $Q_1^* 
= V_c(\vec k_2)Q_1$ and $Q_2^* = V_c(\vec k_1)Q_2$, respectively. The 
corresponding current operator for the OGE interaction may be expressed as
\begin{equation}
\vec\jmath_{\mathrm{ex}}^{\:\:\mathrm{g}} \:\:=\:\: 
-e\,\left(Q_1^*\left[ \frac{i\vec\sigma_1\times\vec 
k_2}{2m_1^2} + \frac{2\vec P_2 + i\vec\sigma_2\times\vec 
k_2}{2m_1m_2}\right] + Q_2^*\left[
\frac{i\vec\sigma_2\times\vec k_1}{2m_2^2} +
\frac{2\vec P_1 + i\vec\sigma_1\times\vec k_1}{2m_1m_2}\right]\right),
\label{Ogecurr}
\end{equation}
with $Q_1^* = V_g(\vec k_2)Q_1$ and $Q_2^* = V_g(\vec k_1)Q_2$. As the 
above equation depends only on $\vec k_1$ and $\vec k_2$, the OGE magnetic 
moment operator is most conveniently calculated using eq.~(\ref{muex2}). 
By Fourier transformation, the resulting magnetic moment operators for 
transitions between $S$-wave quarkonium states may be obtained 
as~\cite{LahdeDs}
\begin{equation}
\vec\mu_{\mathrm{ex}}^{\:\mathrm{Conf}} = 
-\frac{eV_c(r)}{4}\left\{\left[\frac{Q_1}{m_1^2} -
\frac{Q_2}{m_2^2}\right]\:(\vec\sigma_1 - \vec\sigma_2)
+\left[\frac{Q_1}{m_1^2} +
\frac{Q_2}{m_2^2}\right]\:(\vec\sigma_1 + \vec\sigma_2)\right\}
\label{muc}
\end{equation}
for the scalar confining interaction, and
\begin{equation}
\vec\mu_{\mathrm{ex}}^{\:\mathrm{Oge}}
= -\frac{eV_g(r)}{8}\left\{\left[\frac{Q_1}{m_1^2} - \frac{Q_2}{m_2^2}
- \frac{Q_1-Q_2}{m_1m_2}\right]\:(\vec\sigma_1 - \vec\sigma_2)
+\left[\frac{Q_1}{m_1^2} + \frac{Q_2}{m_2^2}
+ \frac{Q_1+Q_2}{m_1m_2}\right]\:(\vec\sigma_1 + \vec\sigma_2)\right\}
\label{mug}
\end{equation}
for the OGE interaction. For equal constituent quark masses, 
eqs.~(\ref{muc}) and~(\ref{mug}) reduce to the expressions given in 
ref.~\cite{Tsushima}. Note that the presence of a spin-flip term in the 
OGE operator~(\ref{mug}) is solely a consequence of the difference in mass 
between the constituent quarks, and will thus not contribute to the M1 
decay widths of the charmonium and bottomonium states. Similarly, the 
terms that are symmetric in the quark spins vanish for equal mass 
quarkonia. However, in the case of the $B_c^\pm$ system, these terms will 
contribute to the magnetic moment of the $c\bar b$ system. Also the 
spin-flip M1 decays in the $B_c^\pm$ system will receive a contribution 
from the OGE operator.

\newpage

\section{Wavefunctions and Decay Widths}

\subsection{Hamiltonian model}
\label{hamsec}

The interaction Hamiltonian employed in this paper for the $Q\bar Q$ 
interaction contains contributions from the scalar confining, 
the vector one-gluon exchange~(OGE), and the instanton induced 
interaction of ref.~\cite{Zahed}. This Hamiltonian is thus of the form
\begin{equation}
H_{\mathrm {int}} = V_{\mathrm {Conf}} + V_{\mathrm {Oge}} + V_{\mathrm
{Inst}}, \label{ham}
\end{equation}
and has been employed together with the covariant Blankenbecler-Sugar 
(BSLT) equation. That equation may be expressed as an eigenvalue equation 
of the form
\begin{equation}
(H_0 + H_{\mathrm {int}})\Psi_{\mathrm{nlm}}(\vec r\,) = 
\varepsilon(E,M_Q,M_{\bar Q})\Psi_{\mathrm{nlm}}(\vec r\,),
\label{blank}
\end{equation}
where $H_0$ denotes the kinetic energy operator of the nonrelativistic 
Schr\"odinger equation and the eigenvalue $\varepsilon(E,M_Q,M_{\bar Q})$ 
is a quadratic mass operator~\cite{2pihq} which is related to the energy 
$E$ of the heavy quarkonium state according to
\begin{equation}
\varepsilon = \frac{\left[ E^2 - (M_Q+M_{\bar Q})^2 \right ]
\left[ E^2 - (M_Q-M_{\bar Q})^2 \right ]}{8\mu E^2} \label{EBSLT}
\end{equation}
where $\mu$ is the reduced mass of the quark-antiquark system. As the 
details of the interaction Hamiltonian~(\ref{ham}) have already been 
described in ref.~\cite{2pihq}, only the main points will be repeated here. 
The interaction operators $V$ in eq.~(\ref{ham}), which in general are 
nonlocal, may be obtained from the $Q\bar Q$ irreducible quasipotential 
$\cal V$ according to
\begin{equation}
V(\vec p\,',\vec p\,) = \sqrt{\frac{M_Q+M_{\bar Q}}{W(\vec p\,')}}\,{\cal
V}(\vec p\,',\vec p\,)\,\sqrt{\frac{M_Q+M_{\bar Q}}{W(\vec p\,)}},
\label{VBSLT}
\end{equation}   
where the function $W$ is defined as $W(\vec p\,) = E_Q(\vec p\,) +
E_{\bar Q}(\vec p\,)$ with $E_i(\vec p\,) = \sqrt{M_i^2 + \vec p\,^2}$. 
Note that in the Born approximation the quasipotential $\cal V$ is set 
equal to the $Q\bar Q$ invariant scattering amplitude $\cal T$, whereby a 
constructive relation to field theory obtains. 

The OGE interaction in momentum space can be parameterized~\cite{Matt} in 
terms of the strong coupling $\alpha_s(k^2)$ according to
\begin{equation}
V_{\mathrm{Oge}}(\vec k\,) = -\frac{16\pi}{3} 
\frac{\alpha_s(\vec k^2)}{\vec k^2}, 
\quad \alpha_s(\vec k^2) = \frac{12\pi}{27}\:\ln^{-1}\left[
\frac{\vec k^2 + 4m_g^2}{\Lambda_{\mathrm{QCD}}^2}\right].
\label{ogefunc}
\end{equation}
Here $\Lambda_{\mathrm{QCD}}$ denotes the QCD scale of $\sim 250$~MeV, and 
$m_g$ is a dynamical gluon mass which determines the low-momentum behavior 
of $\alpha_s$. Application of eq.~(\ref{ogefunc}) together with 
eq.~(\ref{VBSLT}) yields the central and spin-dependent potential 
components of the OGE interaction, and are given in ref.~\cite{2pihq}. 
For example, if all higher order nonlocalities are dropped, the central 
Coulomb component of the OGE interaction is modified to
\begin{equation}
V_{\mathrm {Oge}}\,(r) = - \frac{4}{3}\frac{2}{\pi}
\int_{0}^{\infty}dk\:j_0(kr)\,\frac{M_Q}{e_Q}\frac{M_{\bar Q}}{e_{\bar Q}}
\left(\frac{M_Q+M_{\bar Q}}{e_Q+e_{\bar Q}}\right)\alpha_s(k^2),
\label{OGEpot}
\end{equation}
where the factors $e_Q$ and $e_{\bar Q}$ are defined as
$e_Q=\sqrt{M_Q^2+k^2/4}$ and $e_{\bar Q}=\sqrt{M_{\bar Q}^2+k^2/4}$. If 
$\alpha_s$ is taken to be constant, then the above form reduces to the 
Coulombic potential suggested by perturbative QCD in the limit 
$M_Q\rightarrow\infty$. On the other hand, the Fourier transform of a 
linear confining potential of the form $V_{\mathrm {Conf}} = cr$ may be 
expressed as
\begin{equation}
V_{\mathrm {Conf}}\,(r) = \lim_{\lambda\rightarrow 0} 
\int\frac{d^3k}{(2\pi)^3}\,e^{-i\vec k\cdot\vec r}\:
\frac{8\pi c}{k^4}\left[\frac{4\,\lambda^2k^4}{(\lambda^2 + k^2)^3} - 
\frac{k^4}{(\lambda^2 + k^2)^2}\right].
\end{equation}
For the purpose of calculating the spin-orbit term associated 
with the scalar confining interaction, the limit $\lambda\rightarrow 0$ may 
be taken directly to yield $V_{\mathrm {Conf}}(\vec k\,) = -8\pi c/k^4$. 
Finally the instanton induced interaction is expressed as
\begin{equation}
V_{\mathrm{Inst}}\,(r) = - \frac{\Delta M_Q\Delta M_{\bar Q}}{4n} \:\:
\int_0^\infty dk\:k^2\:j_0(kr)
\left(\frac{M_Q+M_{\bar Q}}{e_Q+e_{\bar Q}}\right)\frac{M_QM_{\bar Q}}{e_Q
e_{\bar Q}},
\label{instpot2}
\end{equation}
where the notation is similar to that employed in ref.~\cite{Zahed}. The 
parameter $\Delta M_Q$ denotes the mass shift of the heavy constituent 
quark due to the instanton induced interaction, which for a charm quark is 
of the order $\sim 100$ MeV~\cite{Zahed}. The parameter $n$ represents the 
instanton density, which is usually taken to be $\sim 1\:\mathrm{fm}^{-4}$. 
Note that in the limit of infinitely heavy constituent quarks, 
eq.~(\ref{instpot2}) reduces to a delta function. The above smeared out 
form is convenient since it allows for direct numerical treatment of the 
instanton induced interaction with the differential equation~(\ref{blank}).

\begin{figure}[h!]
\parbox{.65\textwidth}{
\epsfig{file = 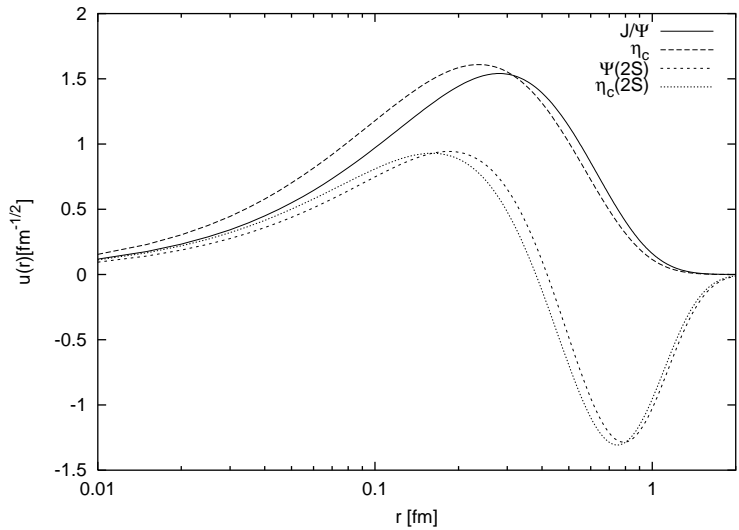}}
\parbox{.34\textwidth}{
\caption{The reduced radial wavefunctions for the charmonium $1S$ and $2S$ 
states, from ref.~\cite{2pihq}. The differences between the spin singlet 
$\eta_c$ and spin triplet $\psi$~\mbox{wavefunctions} are due to the 
short-ranged OGE spin-spin interaction. Note that the $r$ axis has been 
made logarithmic in order to emphasize the short range part of the 
wavefunctions.}
\label{logplot}}
\end{figure}

The wavefunctions needed for the calculation of the E1 and M1 widths of the 
heavy quarkonia are thus taken to be the solutions to eq.~(\ref{blank}) 
obtained in ref.~\cite{2pihq}. As the spin-spin interaction in the 
charmonium system is strong enough to produce a $J/\psi\,-\,\eta_c$ 
splitting of $\sim 120$ MeV, then the respective radial wavefunctions are 
likely to show marked differences. Thus the employment of spin-averaged 
wavefunctions and a perturbative treatment of the spin-dependent hyperfine 
interaction is undesirable. With this in mind, the hyperfine 
components of the $Q\bar Q$ interaction have been taken fully into account, 
which is possible since they are nonsingular in the BSLT framework. In 
particular, it is expected that such effects should be significant for 
transitions which involve a change in the principal quantum number of the 
$Q\bar Q$ state. The effect of the $S$-wave spin-spin interaction on the 
radial wavefunctions is demonstrated by Fig.~\ref{logplot}. The calculated 
spectra of the heavy quarkonia and the model parameters are given in 
Tables~\ref{stat} and~\ref{par}, respectively.

\vspace{.5cm}

\newpage

\begin{table}[h!]
\parbox{0.63\textwidth}{
\begin{tabular}{c||c|c|c|c|c}
$n\,^{2S+1}L_J$ & $b\bar b$ & Exp($b\bar b$) & $c\bar c$ & Exp($c\bar c$)
& $c\bar b$ \\ \hline\hline
&&&&&\\
$1\,^1S_0$ & 9401  &  --  & 2997 & $2980\pm 1.8$ & 6308 \\
$2\,^1S_0$ & 10005 &  --  & 3640 & $3654\pm 6$~\cite{Belle}   & 6888 \\
$3\,^1S_0$ & 10361 &  --  & 4015 &  -- & 7229 \\
$4\,^1S_0$ & 10634 &  --  & 4300 &  -- & 7488 \\
&&&&&\\
$1\,^3S_1$ & 9458  & 9460  & 3099 & 3097 & 6361 \\
$2\,^3S_1$ & 10030 & 10023 & 3678 & 3686 & 6910 \\
$3\,^3S_1$ & 10377 & 10355 & 4040 & $4040\pm 10$ & 7244 \\
$4\,^3S_1$ & 10648 & 10580 & 4319 & $4159\pm 20$ ? & 7500 \\
&&&&&\\
$1\,^1P_1$ & 9888  &  --  & 3513 &  --  & 6754 \\
$2\,^1P_1$ & 10266 &  --  & 3912 &  --  & 7126 \\
$3\,^1P_1$ & 10552 &  --  & 4211 &  --  & 7401 \\
&&&&&\\
$1\,^3P_0$ & 9855  & 9860  & 3464 & 3415 & 6723 \\
$2\,^3P_0$ & 10244 & 10232 & 3884 &  -- & 7107 \\
$3\,^3P_0$ & 10535 &  --   & 4192 &  -- & 7387 \\
&&&&&\\
$1\,^3P_1$ & 9883  & 9893  & 3513 & 3511 & 6751 \\
$2\,^3P_1$ & 10263 & 10255 & 3913 &  -- & 7125 \\
$3\,^3P_1$ & 10550 &  --   & 4213 &  -- & 7400 \\
&&&&&\\
$1\,^3P_2$ & 9903  & 9913  & 3540 & 3556 & 6770 \\
$2\,^3P_2$ & 10277 & 10269 & 3930 &  -- & 7136 \\
$3\,^3P_2$ & 10561 &  --   & 4226 &  -- & 7410 \\
&&&&&\\
$1\,^3D_3$ & 10158 &  --  & 3790 &  --  & 7009 \\ 
$1\,^3D_2$ & 10149 &  --  & 3784 &  --  & 7006 \\ 
$1\,^3D_1$ & 10139 &  --  & 3768 & $3770\pm 2.5$ & 6998 \\ 
\end{tabular}}
\parbox{0.35\textwidth}{
\caption{Calculated and experimental charmonium, bottomonium and $B_c^\pm$ 
states rounded to the nearest MeV. The states are classified according to 
excitation number $n$, total spin $S$, total orbital angular momentum $L$ 
and total angular momentum $J$. Note that experimental uncertainties are 
indicated only where they are appreciable. For a graphical plot of the 
$c\bar c$ and $b\bar b$ data, see \mbox{ref.~\cite{2pihq}.} 
The $c\bar c$ and $b\bar b$ values are 
from ref.~\cite{2pihq}. The experimental \mbox{states} correspond to the 
values 
reported by ref.~\cite{PDG}, except for the recently observed~\cite{Belle} 
$\eta_c(2S)$. The measured mass of the $B_c^\pm$ was reported in 
ref.~\cite{CDF} as $6.40\pm 0.39$ GeV, which is about $\sim 100$ MeV higher 
than the predicted 6308 MeV, and most other models give even lower masses 
for the $B_c^\pm$ ground state~\cite{Quigg}. The predicted $B_c^\pm$ 
spectrum agrees, however, very well with the QCD-inspired model of 
\mbox{ref.~\cite{Chen}.} Within the framework of QCD sum rules, the mass of 
the $B_c^\pm$ has been predicted to be $\sim 6.35$~GeV~\cite{Colangelo}. 
\label{stat}}}
\end{table}

\vspace{.5cm}

\begin{table}[h!]
\parbox{0.50\textwidth}{\centering{
\begin{tabular}{c||c|c}
& Ref.~\cite{2pihq} & Other sources \\ \hline\hline 
&&\\
$M_b$   & 4885 MeV      & 4870 MeV~\cite{Roberts}       \\
$M_c$   & 1500 MeV      & 1530 MeV~\cite{Roberts}       \\ 
\vspace{-.2cm} && \\ 
$\Lambda_{\mathrm{QCD}}$ & 260 MeV & 200-300 MeV~\cite{Matt} \\
$m_g$   & 290 MeV       & $m_g > \Lambda_{\mathrm{QCD}}$~\cite{Matt}\\
$c$     & 890 MeV/fm    & 912 MeV/fm~\cite{Roberts} \\ && \\
$\frac{(\Delta M_c)^2}{4n}$ & 0.084 $\mathrm{fm}^{2}$ & $\sim 0.05$
$\mathrm{fm}^{2}$~\cite{Zahed} \\ 
\vspace{-.2cm} && \\
$\frac{(\Delta M_b)^2}{4n}$ & 0.004 $\mathrm{fm}^{2}$ & -- \\ 
\vspace{-.2cm} && \\
$\frac{\Delta M_b\Delta M_c}{4n}$ & 0.018 $\mathrm{fm}^{2}$ & -- \\
\end{tabular}}}
\parbox{0.48\textwidth}{
\caption{Constituent quark masses and parameters for the $Q\bar Q$ 
interaction that have been used in the calculation of the spectra 
presented in Table~\ref{stat}. The heavy masses are close to those preferred 
by ref.~\cite{Roberts}, and in general in agreement with the values 
of previous work. The form of the running QCD coupling $\alpha_s$ is in 
line with the criteria of refs.~\cite{Matt,Bethke}, and the confining 
string tension~$c$ \mbox{agrees} well with previous 
calculations~\cite{Roberts,Isgur}. The strength of the instanton induced 
interaction for $c\bar c$ is comparable to the estimate given by 
\mbox{ref.~\cite{Zahed}.} }
\label{par}}
\end{table}

\newpage

\subsection{Widths for radiative decay}

The decay width for an E1 dominated transition of the type 
$\chi_{cJ}\rightarrow J/\psi\,\gamma$ or $\psi'\rightarrow 
\chi_{cJ}\,\gamma$ is given by
\begin{equation}
\Gamma\:\:=\:\:{\cal S}_{fi}
\,\frac{2J_f\!+\!1}{3}\:q^3\alpha\:\frac{M_f}{M_i}
\left[\:\frac{4}{9}\:|{\cal M}_0|^2 + \frac{8}{9}\:|{\cal M}_2|^2\right],
\label{dyndec}
\end{equation}
where $J_f$ is the total angular momentum of the final quarkonium state, 
and $q$ is the momentum of the emitted photon. One would thus expect that 
the widths for $\psi'\rightarrow \chi_{cJ}\,\gamma$ with $J = 0,1,2$ would 
scale as $1\,:\,3\,:\,5$ respectively. In 
practice, however, this result is modified by the large hyperfine 
splittings in the $L=1$ heavy quarkonia. The factor ${\cal S}_{fi}$ is 
defined as in ref.~\cite{Quigg} and assumes the values ${\cal S}_{fi} = 1$ 
for a triplet-triplet transition and ${\cal S}_{fi} = 3$ for a 
singlet-singlet transition of the type $h_c\rightarrow \eta_c\gamma$. On 
the other hand, the widths for transitions between $D$- and $P$-wave states 
are calculated according to
\begin{equation}
\Gamma\:\:=\:\:4\,{\cal S}_{fi}
\,\frac{2J_f\!+\!1}{27}\:q^3\alpha\:\frac{M_f}{M_i}\:|{\cal M}_0|^2, 
\quad\quad {\cal S}_{fi}\:\:=\:\: 18\, 
\left\{\begin{array}{ccc}
2   & 1 & J_d \\
J_p & 1 & 1   \\
\end{array}
\right \}^2, \label{Ddec}
\end{equation}
where $J_d$ and $J_p$ are the total angular momenta of the $D$- and 
$P$-wave states, respectively. Note that the triangularity of the 6-j 
symbol requires that $|J_d-J_p|$ = 1 or 0. Consequently, transitions that 
change the value of $J$ by more than one unit are forbidden. In 
eqs.~(\ref{dyndec}) and~(\ref{Ddec}), ${\cal M}_0$ and ${\cal M}_2$ denote 
radial matrix elements for $S$- and $D$-wave photon emission, respectively. 
The radial matrix element for $S$-wave decay receives contributions not 
only from the impulse approximation, eq.~(\ref{cmdip}), but also from the 
confinement and OGE operators~(\ref{cdip}) and~(\ref{gdip}). That matrix 
element may thus be expressed as
\begin{equation}
{\cal M}_0\:\:=\:\:\int_0^\infty dr\:r\,u_f(r)\,u_i(r)\:
j_0\left(\frac{qr}{2}\right)\left[\:\left<Q\right>_{\mathrm{IA}} 
+ q^2\,V_c(r)\,\left<Q\right>_c +
\left(\frac{\partial V_g(r)}{r\,\partial r}\right)
\left<Q\right>_g\right],
\label{smatr}
\end{equation}
where $u_i$ and $u_f$ are the reduced radial wavefunctions for the initial 
and final heavy quarkonium states. Similarly, the matrix element for 
$D$-wave decay is 
of the form
\begin{equation}
{\cal M}_2\:\:=\:\:\left<Q\right>_{\mathrm{ID}}
\int_0^\infty dr\:r\,u_f(r)\,u_i(r)\:
j_2\left(\frac{qr}{2}\right).
\label{dmatr}
\end{equation}
The contribution from this matrix element is generally very small, and has 
therefore not been included in eq.~(\ref{Ddec}).

The impulse approximation charge factor 
$\left<Q\right>_{\mathrm{IA}}$, and the exchange charge factors 
$\left<Q\right>_c$ for the scalar confining interaction and 
$\left<Q\right>_g$ for the OGE interaction that appear in 
eqs.~(\ref{smatr}) and~(\ref{dmatr}) are of the form 
\begin{equation}
\left<Q\right>_{\mathrm{IA}}\:\:=\:\: 
\left[\,Q_1\left(1-\frac{q^2}{8m_1^{\,2}}\right)\,\frac{m_2}{m_1+m_2}\:-\: 
Q_2\left(1-\frac{q^2}{8m_2^{\,2}}\right)\,\frac{m_1}{m_1+m_2}\,\right]
\end{equation}
for the impulse approximation, where the quark charge operators have been 
multiplied with the Darwin-Foldy terms from eq.~(\ref{rsq}), and
\begin{eqnarray}
\left<Q\right>_c &=& 
\left[\frac{Q_1}{4m_1^{\,3}}\,\frac{m_2}{m_1+m_2} -
\frac{Q_2}{4m_2^{\,3}}\,\frac{m_1}{m_1+m_2}\right], \\
\left<Q\right>_g &=& 
\left[\frac{Q_1}{4m_1^{\,2}}\left(\frac{1}{m_1} +
\frac{2}{3}\frac{\left<S_f|\vec\sigma_1\!\cdot\vec\sigma_2|S_i\right>}
{m_2}\right) -\frac{Q_2}{4m_2^{\,2}}\left(\frac{1}{m_2} +
\frac{2}{3}\frac{\left<S_f|\vec\sigma_1\!\cdot\vec\sigma_2|S_i\right>}
{m_1}\right)\right],\label{OGEQ}
\end{eqnarray}
for the confinement and OGE exchange charge contributions, respectively. In 
the spin dependent terms of eq.~(\ref{OGEQ}), $S_i$ and $S_f$ denote the 
total spins of the initial and final quarkonium states. For triplet-triplet 
and singlet-singlet transitions,  
$\left<S_f|\vec\sigma_1\!\cdot\vec\sigma_2|S_i\right> = +1$ and $-3$, 
respectively. The charge factor $\left<Q\right>_{\mathrm{ID}}$ that 
appears in eq.~(\ref{dmatr}) is defined according to 
$\left<Q\right>_{\mathrm{ID}} = \lim_{q\rightarrow 
0}\left<Q\right>_{\mathrm{IA}}$. This is permissible since the Darwin-Foldy 
and exchange charge terms are very small compared to the dominant dipole 
contribution, which in itself is already insignificant because of the 
suppression by the $j_2$ function in the matrix element. There is thus no 
need to include these terms in the matrix element ${\cal M}_2$ for $D$-wave 
decay. 

In the E1 
approximation, the recoil factor $M_f/M_i$ vanishes by eq.~(\ref{matr2.5}), 
since in that case $\vec P_f = \vec P_i$. The decay width 
expression~(\ref{dyndec}) thus reduces to
\begin{equation}
\Gamma_{\mathrm{E1}}\:=\:\:{\cal S}_{fi}\frac{2J_f\!+\!1}{27}\:
4\:q_{\mathrm{nr}}^3\:\alpha\:
|\lim_{q\rightarrow 0}{\cal M}_0|^2,
\label{E1dec}
\end{equation}
in the E1 approximation, which is similar to the expression given in 
ref.~\cite{Quigg}. Here the "recoilless" \mbox{$q$-value} is given by 
$q_{\mathrm{nr}} = M_i-M_f$. Note that the OGE exchange charge contribution 
survives in the E1 approximation, whereas the contribution from the scalar 
confining interaction as well as the Darwin-Foldy terms are eliminated.

The expression for the width of a spin-flip M1 transition between 
$S$-wave heavy quarkonium states can be written in the form
\begin{equation}
\Gamma_{\mathrm{M1}}\:=\:\:\frac{16}{2S_i\!+\!1}\:q^3\alpha\:\frac{M_f}{M_i}
\:|{\cal M}_{\gamma}|^2,
\label{M1dec}
\end{equation}
where ${\cal M}_{\gamma}$ denotes the radial matrix element for M1 decay 
and $S_i$ is the total spin of the initial state. The radial matrix 
element for M1 decay consists of relativistic impulse approximation, scalar 
confining and OGE components, according to
\begin{equation}
{\cal M}_{\gamma}\:\:=\:\:{\cal M}_{\gamma}^{\mathrm{RIA}} + {\cal 
M}_{\gamma}^{\mathrm{Conf}} + {\cal M}_{\gamma}^{\mathrm{Oge}},
\label{M1matr}
\end{equation}
where the different matrix elements can be obtained from the corresponding 
spin-flip magnetic moment operators by dropping the charge $e$ and the 
spinors $\vec\sigma_1-\vec\sigma_2$. From the relativistic matrix 
element~(\ref{matr4}) and the relativistic spin-flip magnetic moment 
operator~(\ref{RIA}), one obtains
\begin{equation}
{\cal M}_{\gamma}^{\mathrm{RIA}}\:\:=\:\:
\frac{2}{\pi} \int_0^\infty dr'\,r'u_f(r') \int_0^\infty dr\,r\,u_i(r) 
\int_0^\infty dP\,P^2 \: \frac{1}{4}\left[\frac{Q_1}{m_1}f_1^\gamma -
\frac{Q_2}{m_2}f_2^\gamma\right]\: j_0\left(r'P\right)j_0\left(rP\right),
\label{RIAmatr}
\end{equation}
where the factors $f_i^\gamma$ are given by eq.~(\ref{RIA}). The matrix 
elements associated with the scalar confining and vector OGE interactions 
are of the form
\begin{equation}
{\cal M}_\gamma^{\mathrm{Conf}}\:\:=\:\: -\int_0^\infty
dr\,u_f(r)\,u_i(r)\,\frac{V_c(r)}{4}\left[\frac{Q_1}{m_1^{\,2}} -
\frac{Q_2}{m_2^{\,2}}\right],
\label{NRConf}
\end{equation}  
for the scalar confining interaction, and
\begin{equation}
{\cal M}_\gamma^{\mathrm{Oge}}\:\:=\:\: -\int_0^\infty
dr\,u_f(r)\,u_i(r)\,\frac{V_g(r)}{8}\left[\frac{Q_1}{m_1^{\,2}} -
\frac{Q_2}{m_2^{\,2}} - \frac{Q_1-Q_2}{m_1 m_2}\right],
\label{NROge}
\end{equation}
for the OGE interaction. 

It should be noted at this point that the 
appearance of two-quark matrix elements as given above is entirely due to 
the elimination in the impulse approximation of the negative energy 
components in the reduction of the Bethe-Salpeter equation to a 
Blankenbecler-Sugar equation. These components do however contribute as 
transition matrix elements, and have to be included in the 
Blankenbecler-Sugar (or Schr\"odinger) framework as explicit two-quark 
current operators~\cite{Coester}, as illustrated in Fig.~\ref{feyn}. In 
e.g. the alternate Gross type reduction of the Bethe-Salpeter equation, 
these two-quark operators are automatically taken into account by the 
single quark transition operators.

\newpage

\subsection{Magnetic moment of the $B_c^\pm$}

Of the heavy quarkonium systems, only the spin-triplet $B_c^{\pm}$ ($c\bar 
b$, $\bar c b$) states have a magnetic moment. That may be calculated from 
the spin-symmetric part of eq.~(\ref{nrscurr}). When expressed in terms of 
the nuclear magneton~$\mu_N$, the magnetic moment of the $B_c^{*+}$~($c\bar 
b$) state is obtained as
\begin{equation}
\mu_{c\bar b}\:=\:\frac{m_p}{3}\left[\frac{2}{m_c} + \frac{1}{m_b}\right]
\mu_N, \label{NRmu}
\end{equation}
where $m_p$ is the proton mass. The above expression is valid in the 
nonrelativistic limit. Since it is known that the magnetic moments of the 
baryons receive significant corrections from relativistic 
effects~\cite{Dannbom}, then a relativistic version of eq.~(\ref{NRmu}) is 
called for. In the relativistic impulse approximation, the magnetic moment 
of an $S$-wave $c\bar b$ state is of the form
\begin{equation}
\mu_{c\bar b}^{\mathrm{RIA}}\:=\:\frac{2}{\pi}\int_0^\infty dr\,dr'\,r\,
r'\,u(r)\,u(r') \int_0^\infty dP\,P^2 \: 
\frac{m_p}{3}\left[\frac{2}{m_c}f_c^\gamma + 
\frac{1}{m_b}f_b^\gamma\right]\:j_0\left(r'P\right)j_0\left(rP\right)\:
\mu_N,\label{RIAmu}
\end{equation}
where $u(r)$ is the reduced radial wavefunction of the $c\bar b$ state. In 
eq.~(\ref{RIAmu}), the factors $f_i^\gamma$ are defined in 
eq.~(\ref{RIA}). In addition to the relativistic impulse approximation, 
the exchange magnetic moment operators associated with the scalar confining 
and vector OGE interactions will also contribute to the magnetic moment of 
the $B_c^\pm$. The contribution from the scalar confining interaction is of 
the form
\begin{equation}
\mu_{c\bar b}^{\mathrm{Conf}}\:=\:-\int_0^\infty dr\,u^2(r)\,V_c(r)\:
\frac{m_p}{3}\left[\frac{2}{m_c^{\,2}} + \frac{1}{m_b^{\,2}}\right]\:
\mu_N, \label{Confmu}
\end{equation}
where the confining interaction $V_c(r)$ is of the form $V_c(r) = cr$. 
Similarly, the OGE contribution can be expressed as
\begin{equation}
\mu_{c\bar b}^{\mathrm{Oge}}\:=\:-\int_0^\infty dr\,u^2(r)\,V_g(r)\:
\frac{m_p}{6}\left[\frac{2}{m_c^{\,2}} + \frac{1}{m_b^{\,2}}
+ \frac{3}{m_cm_b}\right]\:
\mu_N, \label{OGEmu}
\end{equation}
where the OGE potential $V_g(r)$ is given by the Fourier transform of 
eq.~(\ref{ogefunc}). The total magnetic moment of the $B_c^{*\pm}$ is thus 
given by the sum of the RIA contribution~(\ref{RIAmu}) and the exchange 
contributions~(\ref{Confmu}) and~(\ref{OGEmu}).

\section{Numerical Results}

This section presents the numerical values of the widths and matrix 
elements for each E1 and M1 transition, as obtained using the model 
for the $Q\bar Q$ spectra presented in section~\ref{hamsec}. Unless 
otherwise indicated, the E1 widths have been calculated using 
eqs.~(\ref{dyndec}),~(\ref{Ddec}),~(\ref{smatr}) and~(\ref{dmatr}), while 
the widths for M1 decay correspond to eqs.~(\ref{M1dec}) 
and~(\ref{M1matr}). When the mass of one of the quarkonium states is not 
known empirically, then the splittings rather than the absolute values from 
Table~\ref{stat} are used to determine the photon momentum $q_\gamma$.

The widths for M1 transitions between $S$-wave states in charmonium and 
bottomonium are given in Table~\ref{M1tab}, along with the associated 
impulse approximation and exchange current matrix elements. For $c\bar c$ 
and $b\bar b$, the width for M1 decay is given by the sum of the impulse 
approximation and scalar confinement terms. In the case of the $B_c^\pm$, the 
widths for M1 decay also receive a contribution from the OGE exchange current.
The M1 widths and magnetic moments of the $B_c^\pm$ states are given in 
Tables~\ref{M1tabBc} and~\ref{Mtab}, respectively. The widths of the E1 
dominated transitions between low-lying $c\bar c$ and $b\bar b$ states are 
given in Tables~\ref{ccE1},~\ref{bbE1},~\ref{E1Dcc},~\ref{E1Dbb} 
and~\ref{E1rest}. They have been calculated both for the dynamical model 
presented in section~\ref{E1op} and the rigorous E1 approximation. The 
computed widths for E1 dominated transitions in the $B_c^\pm$ system are 
given, along with the matrix elements, in Tables~\ref{bcE1},~\ref{bcrest} 
and~\ref{E1Dbc}. 

\newpage

\begin{table}[h!]
\begin{center}
\caption{The M1 transitions between low-lying $S$-wave states in the 
charmonium~($c\bar c$) and bottomonium~($b\bar b$) systems. Experimental 
data~\cite{PDG} is available only for the $J/\psi\rightarrow \eta_c\gamma$ 
and $\psi\,'\rightarrow \eta_c\gamma$ transitions. Note that the empirical 
value for $\psi\,'\rightarrow \eta_c\gamma$ is uncertain since the total 
width of the $\psi\,'$ is poorly known. The quoted photon momenta 
$q_\gamma$ have been obtained by combination of the empirical masses of the 
spin triplet states~\cite{PDG} with the splittings given by the Hamiltonian 
model of ref.~\cite{2pihq} in Table~\ref{stat}. The M1 decays of 
the $\psi(3S)$ state have not been included since the $3S$ states in 
charmonium lie above the threshold for $D\bar D$ fragmentation.}
\label{M1tab}
\vspace{.5cm}
\begin{tabular}{l||r|r|r||c|c|c}
\multicolumn{1}{c||}{Transition} & 
\multicolumn{3}{c||}{Matrix element\,\,[fm]} & \multicolumn{3}{c}{Width} \\
&\multicolumn{3}{c||}{}&\multicolumn{3}{c}{} \\
& \multicolumn{1}{c|}{NRIA} & \multicolumn{1}{c|}{RIA} & 
\multicolumn{1}{c||}{Conf} & \,NRIA\, & \,RIA\, & {\bf RIA+Conf} \\ 
\hline\hline
&&&&&& \\
$J/\psi\rightarrow \eta_c\gamma$ & $4.356\cdot 10^{-2}$ & 
$3.762\cdot 10^{-2}$ & $-8.724\cdot 10^{-3}$ & 2.85 & 2.12 & {\bf 1.25 keV} \\
\vspace{-.2cm}
\footnotesize{$q_\gamma : 116$ MeV}&&&&&& \footnotesize{exp: $1.14\pm 
0.39$} \\
&&&&&& \\
$\psi{\,'}\rightarrow \eta_c\gamma$  & $3.985\cdot 10^{-3}$ & 
$-5.14\cdot 10^{-4}$ & $ 2.826\cdot 10^{-3}$ & 3.35 & 0.06 & {\bf 1.13 keV} \\
\vspace{-.2cm}
\footnotesize{$q_\gamma : 639$ MeV}&&&&&& \footnotesize{exp: $0.84\pm 
0.24$} \\
&&&&&& \\
$\psi{\,'}\rightarrow \eta_c'\gamma$ & $4.344\cdot 10^{-2}$ & 
$3.735\cdot 10^{-2}$ & $-1.870\cdot 10^{-2}$ & 0.18 & 0.13 & {\bf 0.03 keV} \\
\vspace{-.2cm}
\footnotesize{$q_\gamma : 46$ MeV}&&&&&& \\
&&&&&& \\
$\eta_c'\rightarrow J/\psi\,\gamma$ & $-4.271\cdot 10^{-3}$ & 
$-7.584\cdot 10^{-3}$ & $5.206\cdot 10^{-3}$ & 5.89 & 18.6 & {\bf 1.83 keV} \\
\vspace{-.2cm}
\footnotesize{$q_\gamma : 502$ MeV}&&&&&& \\
&&&&&& \\
\hline
&&&&&& \\
$\Upsilon\rightarrow \eta_b\gamma$ & $-6.71\cdot 10^{-3}$ & 
$-6.39\cdot 10^{-3}$ & $ 2.41\cdot 10^{-4}$ & 9.2 & 8.3 & {\bf 7.7 eV} \\
\vspace{-.2cm}
\footnotesize{$q_\gamma : 59$ MeV}&&&&&& \\
&&&&&& \\
$\Upsilon(2S)\rightarrow \eta_b\gamma$ & $-3.94\cdot 10^{-4}$ & 
$-1.44\cdot 10^{-4}$ & $ -8.76\cdot 10^{-5}$ & 31.8 & 4.3 & {\bf 11.0 eV} \\
\vspace{-.2cm}
\footnotesize{$q_\gamma : 603$ MeV}&&&&&& \\
&&&&&& \\
$\Upsilon(2S)\rightarrow \eta_b(2S)\,\gamma$ & $-6.70\cdot 10^{-3}$ & 
$-6.39\cdot 10^{-3}$ & $ 5.45\cdot 10^{-4}$ & 0.70 & 0.64 & {\bf 0.53 eV} \\
\vspace{-.2cm}
\footnotesize{$q_\gamma : 25$ MeV}&&&&&& \\
&&&&&& \\
$\eta_b(2S)\rightarrow \Upsilon\,\gamma$ & $4.18\cdot 10^{-4}$ & 
$6.30\cdot 10^{-4}$ & $ -1.31\cdot 10^{-4}$ & 71.5 & 162 & {\bf 102 eV} \\
\vspace{-.2cm}
\footnotesize{$q_\gamma : 530$ MeV}&&&&&& \\
&&&&&& \\
$\Upsilon(3S)\rightarrow \eta_b(3S)\,\gamma$ & $-6.70\cdot 10^{-3}$ & 
$-6.35\cdot 10^{-3}$ & $ 8.02\cdot 10^{-4}$ & 0.18 & 0.16 & {\bf 0.13 eV} \\
\vspace{-.2cm}
\footnotesize{$q_\gamma : 16$ MeV}&&&&&& \\
&&&&&& \\
$\Upsilon(3S)\rightarrow \eta_b(2S)\,\gamma$ & $-3.59\cdot 10^{-4}$ & 
$-1.11\cdot 10^{-4}$ & $ -1.55\cdot 10^{-4}$ & 5.3 & 0.5 & {\bf 2.9 eV} \\
\vspace{-.2cm}
\footnotesize{$q_\gamma : 350$ MeV}&&&&&& \\
&&&&&& \\
$\Upsilon(3S)\rightarrow \eta_b\,\gamma$ & $-2.10\cdot 10^{-4}$ & 
$-6.59\cdot 10^{-5}$ & $ -3.77\cdot 10^{-5}$ & 30.2 & 3.0 & {\bf 7.3 eV} \\
\vspace{-.2cm}
\footnotesize{$q_\gamma : 910$ MeV}&&&&&& \\
&&&&&& \\
$\eta_b(3S)\rightarrow \Upsilon(2S)\,\gamma$ & $3.96\cdot 10^{-4}$ & 
$6.05\cdot 10^{-4}$ & $ -2.25\cdot 10^{-4}$ & 13.7 & 32.0 & {\bf 12.6 eV} \\
\vspace{-.2cm}
\footnotesize{$q_\gamma : 311$ MeV}&&&&&& \\
&&&&&& \\
$\eta_b(3S)\rightarrow \Upsilon\gamma$ & $2.05\cdot 10^{-4}$ & 
$3.02\cdot 10^{-4}$ & $ -4.68\cdot 10^{-5}$ & 68.7 & 149 & {\bf 106 eV} 
\\
\vspace{-.2cm}
\footnotesize{$q_\gamma : 842$ MeV}&&&&&& \\
\end{tabular}
\end{center}
\end{table}

\newpage

\begin{table}[h!]
\centering{
\caption{The E1 dominated transitions between low-lying states in the 
charmonium~($c\bar c$) system, together with the empirical data given by 
ref.~\cite{Hagiwara}. The column "IA" contains the matrix 
element~(\ref{smatr}) in the impulse (single quark) approximation, while in 
the column labeled "DYN", the exchange charge contributions have been 
included. The columns "E1" contain the matrix element and the $\gamma$ 
width in the E1 approximation. The $q_\gamma$ values, as given above, have 
been rounded to the nearest MeV, and correspond wherever possible to the 
empirical data in Table~\ref{stat}. Further E1 transitions can be found in 
Table~\ref{E1rest}.} 
\label{ccE1}
\vspace{.4cm}
\begin{tabular}{l||r|r|r|r||c|c}
\multicolumn{1}{c||}{Transition} &
\multicolumn{3}{c|}{${\cal M}_0$\,\,[fm]} & 
\multicolumn{1}{c||}{${\cal M}_2$\,\,[fm]}
&\multicolumn{2}{c}{Width} \\
&\multicolumn{3}{c|}{}& &\multicolumn{2}{c}{} \\
&\multicolumn{1}{c|}{IA} & \multicolumn{1}{c|}{DYN} &
\multicolumn{1}{c|}{E1} & \,\, & \,E1\, & {\bf DYN} \\
\hline\hline
&&&&&& \\
$\chi_{c2}\rightarrow J/\psi\,\gamma$ & \,\,0.2389\,\, & \,\,0.2442\,\, & 
\,\,0.2632\,\, & $7.145\cdot 10^{-3}$ & 558 keV & {\bf 343 keV} \\
\vspace{-.2cm}
\footnotesize{$q_\gamma : 429$ MeV}&&&&&& 
\footnotesize{exp: $389\pm 60$} \\
&&&&&& \\
$\chi_{c1}\rightarrow J/\psi\,\gamma$ & \,\,0.2464\,\, & \,\,0.2519\,\, & 
\,\,0.2673\,\, & $5.729\cdot 10^{-3}$ & 422 keV & {\bf 276 keV} \\
\vspace{-.2cm}
\footnotesize{$q_\gamma : 390$ MeV}&&&&&& 
\footnotesize{exp: $290\pm 60$} \\
&&&&&& \\
$\chi_{c0}\rightarrow J/\psi\,\gamma$ & \,\,0.2556\,\, & \,\,0.2612\,\, & 
\,\,0.2701\,\, & $3.345\cdot 10^{-3}$ & 196 keV & {\bf 144 keV} \\
\vspace{-.2cm}
\footnotesize{$q_\gamma : 303$ MeV}&&&&&& 
\footnotesize{exp: $165\pm 40$} \\
&&&&&& \\
$\psi\,'\rightarrow \chi_{c0}\,\gamma$ & \,\,$-0.2685$\,\, & 
\,\,$-0.2686$\,\, & 
\,\,$-0.2840$\,\, & $-6.106\cdot 10^{-3}$ & 44.6 keV & {\bf 33.1 keV} \\
\vspace{-.2cm}
\footnotesize{$q_\gamma : 261$ MeV}&&&&&& 
\footnotesize{exp: $26.1\pm 4.5$} \\
&&&&&& \\
$\psi\,'\rightarrow \chi_{c1}\,\gamma$ & \,\,$-0.3126$\,\, & 
\,\,$-0.3126$\,\, & 
\,\,$-0.3202$\,\, & $-3.028\cdot 10^{-3}$ & 45.8 keV & {\bf 38.7 keV} \\
\vspace{-.2cm}
\footnotesize{$q_\gamma : 171$ MeV}&&&&&& 
\footnotesize{exp: $25.2\pm 4.5$} \\
&&&&&& \\
$\psi\,'\rightarrow \chi_{c2}\,\gamma$ & \,\,$-0.3440$\,\, & 
\,\,$-0.3442$\,\, & 
\,\,$-0.3489$\,\, & $-1.871\cdot 10^{-3}$ & 37.1 keV & {\bf 33.1 keV} \\
\vspace{-.2cm}
\footnotesize{$q_\gamma : 127$ MeV}&&&&&& 
\footnotesize{exp: $20.4\pm 4.0$} \\
&&&&&& \\
$h_c\rightarrow \eta_c\,\gamma$ & \,\,0.2098\,\, & \,\,0.2091\,\,  & 
\,\,0.2289\,\, & $7.377\cdot 10^{-3}$ & 661 keV & {\bf 370 keV}  \\
\vspace{-.2cm}
\footnotesize{$q_\gamma : 493$ MeV $\:\:{\cal S}_{fi} : 3 $}&&&&&& \\
&&&&&& \\
$\eta_c'\rightarrow h_c\,\gamma$ & \,\,$-0.3420$\,\, & \,\,$-0.3424$\,\, 
& \,\,$-0.3465$\,\, & $-1.618\cdot 10^{-3}$ & 61.5 keV & {\bf 55.0 keV} \\
\vspace{-.2cm}
\footnotesize{$q_\gamma : 125$ MeV $\:\:{\cal S}_{fi} : 3 $}&&&&&& \\
&&&&&& \\
\hline
&&&&&& \\
$\psi(3S)\rightarrow \chi_{c0}\,\gamma$ & \,\,$-0.0456$\,\, & \,\,
$-0.0450$\,\, & \,\,$-0.0199$\,\, & $0.926\cdot 10^{-2}$ & 2.69 keV & 
{\bf 9.86 keV}  \\
\vspace{-.2cm}
\footnotesize{$q_\gamma : 577$ MeV}&&&&&& \\
&&&&&& \\
$\psi(3S)\rightarrow \chi_{c1}\,\gamma$ & \,\,$-0.0306$\,\, & \,\,   
$-0.0298$\,\, & \,\,$-0.0033$\,\, & $1.016\cdot 10^{-2}$ & 0.13 keV &   
{\bf 9.57 keV} \\
\vspace{-.2cm}
\footnotesize{$q_\gamma : 494$ MeV}&&&&&& \\
&&&&&& \\
$\psi(3S)\rightarrow \chi_{c2}\,\gamma$ & \,\,$-0.0168$\,\, & \,\,   
$-0.0161$\,\, & \,\,0.0123\,\, & $1.099\cdot 10^{-2}$ & 2.38 keV &   
{\bf 5.75 keV} \\
\vspace{-.2cm}
\footnotesize{$q_\gamma : 455$ MeV}&&&&&& \\
&&&&&& \\
$\psi(3S)\rightarrow \chi_{c0}(2P)\,\gamma$ & \,\,$-0.4315$\,\, & \,\,   
$-0.4315$\,\, & \,\,$-0.4497$\,\, & $-7.344\cdot 10^{-3}$ & 21.3 keV &   
{\bf 17.8 keV} \\
\vspace{-.2cm}
\footnotesize{$q_\gamma : 153$ MeV}&&&&&& \\
&&&&&& \\
$\psi(3S)\rightarrow \chi_{c1}(2P)\,\gamma$ & \,\,$-0.4860$\,\, & \,\,   
$-0.4861$\,\, & \,\,$-0.4995$\,\, & $-5.399\cdot 10^{-3}$ & 42.6 keV &   
{\bf 37.3 keV} \\
\vspace{-.2cm}
\footnotesize{$q_\gamma : 125$ MeV}&&&&&& \\
&&&&&& \\
$\psi(3S)\rightarrow \chi_{c2}(2P)\,\gamma$ & \,\,$-0.5280$\,\, & \,\,   
$-0.5283$\,\, & \,\,$-0.5391$\,\, & $-4.367\cdot 10^{-3}$ & 53.7 keV &   
{\bf 48.2 keV} \\
\vspace{-.2cm}
\footnotesize{$q_\gamma : 109$ MeV}&&&&&& \\
\end{tabular}}
\end{table}

\newpage

\begin{table}[h!]
\centering{
\caption{The E1 dominated transitions between low-lying states in the 
bottomonium~($b\bar b$) system, together with the empirical data given by 
ref.~\cite{PDG}. The column "IA" contains the matrix element~(\ref{smatr}) 
in the impulse (single quark) approximation, while in the column labeled 
"DYN", the exchange charge contributions have been included. The columns 
"E1" contain the matrix element and the $\gamma$ width in the E1 
approximation. The photon momenta $q_\gamma$ have been obtained as for 
Table~\ref{ccE1}. Further E1 transitions that involve the $\chi_{bJ}(2P)$ 
states can be found in Table~\ref{E1rest}.} 
\label{bbE1}
\vspace{.4cm}
\begin{tabular}{l||r|r|r|r||c|c}
\multicolumn{1}{c||}{Transition} &
\multicolumn{3}{c|}{${\cal M}_0$\,\,[fm]} & 
\multicolumn{1}{c||}{${\cal M}_2$\,\,[fm]}
&\multicolumn{2}{c}{Width} \\
&\multicolumn{3}{c|}{}& &\multicolumn{2}{c}{} \\
&\multicolumn{1}{c|}{IA} & \multicolumn{1}{c|}{DYN} &
\multicolumn{1}{c|}{E1} & \,\, & \,E1\, & {\bf DYN} \\
\hline\hline
&&&&&& \\
$\chi_{b2}\rightarrow \Upsilon\,\gamma$ & \,\,$-0.0721$\,\, & 
\,\,$-0.0723$\,\, & \,\,$-0.0743$\,\, & $-7.76\cdot 10^{-4}$ 
& 42.7 keV & {\bf 36.0 keV} \\
\vspace{-.2cm}
\footnotesize{$q_\gamma : 443$ MeV}&&&&&& 
\footnotesize{Br: $22\pm 3\:\%$} \\
&&&&&& \\
$\chi_{b1}\rightarrow \Upsilon\,\gamma$ & \,\,$-0.0731$\,\, & 
\,\,$-0.0733$\,\, & \,\,$-0.0751$\,\, & $-6.96\cdot 10^{-4}$ 
& 38.1 keV & {\bf 32.5 keV} \\
\vspace{-.2cm}
\footnotesize{$q_\gamma : 424$ MeV}&&&&&& 
\footnotesize{Br: $35\pm 8\:\%$} \\
&&&&&& \\
$\chi_{b0}\rightarrow \Upsilon\,\gamma$ & \,\,$-0.0742$\,\, & 
\,\,$-0.0743$\,\, & \,\,$-0.0759$\,\, & $-5.80\cdot 10^{-4}$ 
& 30.7 keV & {\bf 26.6 keV} \\
\vspace{-.2cm}
\footnotesize{$q_\gamma : 392$ MeV}&&&&&& 
\footnotesize{Br: $<\,6\:\%$} \\
&&&&&& \\
$\Upsilon(2S)\rightarrow \chi_{b0}\,\gamma$ & \,\,$0.0935$\,\, & 
\,\,$0.0935$\,\, & \,\,$0.0942$\,\, & $2.90\cdot 10^{-4}$ 
& 1.07 keV & {\bf 1.01 keV} \\
\vspace{-.2cm}
\footnotesize{$q_\gamma : 162$ MeV}&&&&&& 
\footnotesize{exp: $1.7\pm 0.5$} \\
&&&&&& \\
$\Upsilon(2S)\rightarrow \chi_{b1}\,\gamma$ & \,\,$0.1007$\,\, & 
\,\,$0.1007$\,\, & \,\,$0.1012$\,\, & $2.01\cdot 10^{-4}$ 
& 1.88 keV & {\bf 1.80 keV} \\
\vspace{-.2cm}
\footnotesize{$q_\gamma : 129$ MeV}&&&&&& 
\footnotesize{exp: $3.0\pm 0.7$} \\
&&&&&& \\
$\Upsilon(2S)\rightarrow \chi_{b2}\,\gamma$ & \,\,$0.1063$\,\, & 
\,\,$0.1063$\,\, & \,\,$0.1067$\,\, & $1.53\cdot 10^{-4}$ 
& 2.10 keV & {\bf 2.03 keV} \\
\vspace{-.2cm}
\footnotesize{$q_\gamma : 109$ MeV}&&&&&& 
\footnotesize{exp: $3.1\pm 0.7$} \\
&&&&&& \\
$h_b\rightarrow \eta_b\,\gamma$ & \,\,$-0.0664$\,\, & \,\,$-0.0664$\,\,  & 
\,\,$-0.0685$\,\, & $-7.98\cdot 10^{-4}$ 
& 47.9 keV & {\bf 39.7 keV}  \\
\vspace{-.2cm}
\footnotesize{$q_\gamma : 485$ MeV $\:\:{\cal S}_{fi} : 3 $}&&&&&& \\
&&&&&& \\
$\eta_b(2S)\rightarrow h_b\,\gamma$ & \,\,$0.1066$\,\, & \,\,$0.1066$\,\, 
& \,\,$0.1069$\,\, & $1.20\cdot 10^{-4}$ 
& 2.86 keV & {\bf 2.77 keV}   \\
\vspace{-.2cm}
\footnotesize{$q_\gamma : 100$ MeV $\:\:{\cal S}_{fi} : 3 $}&&&&&& \\
&&&&&& \\
\hline
&&&&&& \\
$\Upsilon(3S)\rightarrow \chi_{b0}\,\gamma$ & \,\,$0.0068$\,\, & \,\,
$0.0068$\,\, & \,\,$0.0039$\,\, & $-1.13\cdot 10^{-3}$ & 0.05 keV & 
{\bf 0.15 keV}  \\
\vspace{-.2cm}
\footnotesize{$q_\gamma : 483$ MeV}&&&&&& \\
&&&&&& \\
$\Upsilon(3S)\rightarrow \chi_{b1}\,\gamma$ & \,\,$0.0036$\,\, & \,\,   
$0.0036$\,\, & \,\,$0.0005$\,\, & $-1.20\cdot 10^{-3}$ & 2.2 {\bf eV} &   
{\bf 0.11 keV} \\
\vspace{-.2cm}
\footnotesize{$q_\gamma : 452$ MeV}&&&&&& \\
&&&&&& \\
$\Upsilon(3S)\rightarrow \chi_{b2}\,\gamma$ & \,\,$0.0007$\,\, & \,\,   
$0.0007$\,\, & \,\,$-0.0025$\,\, & $-1.26\cdot 10^{-3}$ & 0.08 keV &   
{\bf 0.04 keV} \\
\vspace{-.2cm}
\footnotesize{$q_\gamma : 433$ MeV}&&&&&& \\
&&&&&& \\
$\Upsilon(3S)\rightarrow \chi_{b0}(2P)\,\gamma$ & \,\,$0.1505$\,\, & \,\,   
$0.1504$\,\, & \,\,$0.1520$\,\, & $6.18\cdot 10^{-4}$ & 1.19 keV &   
{\bf 1.14 keV} \\
\vspace{-.2cm}
\footnotesize{$q_\gamma : 122$ MeV}&&&&&& 
\footnotesize{exp: $1.4\pm 0.3$} \\
&&&&&& \\
$\Upsilon(3S)\rightarrow \chi_{b1}(2P)\,\gamma$ & \,\,$0.1614$\,\, & \,\,   
$0.1613$\,\, & \,\,$0.1624$\,\, & $4.36\cdot 10^{-4}$ & 2.20 keV &   
{\bf 2.12 keV} \\
\vspace{-.2cm}
\footnotesize{$q_\gamma : 100$ MeV}&&&&&& 
\footnotesize{exp: $3.0\pm 0.5$} \\
&&&&&& \\
$\Upsilon(3S)\rightarrow \chi_{b2}(2P)\,\gamma$ & \,\,$0.1699$\,\, & \,\,   
$0.1699$\,\, & \,\,$0.1707$\,\, & $3.38\cdot 10^{-4}$ & 2.57 keV &   
{\bf 2.50 keV} \\
\vspace{-.2cm}
\footnotesize{$q_\gamma : 86$ MeV}&&&&&& 
\footnotesize{exp: $3.0\pm 0.6$} \\
\end{tabular}}
\end{table}

\newpage

\begin{table}[h!]
\centering{
\caption{The E1 transitions from the lightest spin-triplet $D$-wave states
in charmonium~($c\bar c$). The labeling of the columns is as for 
Tables~\ref{ccE1} and~\ref{bbE1}, and the photon momenta $q_\gamma$ have 
been calculated from the \mbox{$D$-wave} masses predicted in Table~\ref{stat}. 
The statistical factors ${\cal S}_{fi}$ are given by eq.~(\ref{Ddec}). Note 
that a good experimental candidate~\cite{PDG} for the $^3D_1$ state is the 
$\psi\,(3770)$ resonance.}
\label{E1Dcc}
\vspace{.4cm}
\begin{tabular}{l||r|r|r||c|c}
\multicolumn{1}{c||}{Transition} &
\multicolumn{3}{c||}{${\cal M}_0$\,\,[fm]} & \multicolumn{2}{c}{Width} \\
&\multicolumn{3}{c||}{} & \multicolumn{2}{c}{} \\
&\multicolumn{1}{c|}{IA} & \multicolumn{1}{c|}{DYN} &
\multicolumn{1}{c||}{E1} & \,E1\, & {\bf DYN} \\
\hline\hline
&&&&& \\
$^3D_3 \rightarrow \chi_{c2}\,\gamma$ & \,\,$0.4164$\,\, & 
\,\,$0.4194$\,\, & \,\,$0.4353$\,\, & 243 keV & {\bf 192 keV} \\
\vspace{-.22cm}
\footnotesize{$q_\gamma : 227$ MeV $\:\:{\cal S}_{fi} : 18/25 $}&&&&& \\
&&&&& \\
$^3D_2 \rightarrow \chi_{c2}\,\gamma$ & \,\,$0.4188$\,\, & 
\,\,$0.4219$\,\, & \,\,$0.4367$\,\, & 56.5 keV & {\bf 45.2 keV} \\
\vspace{-.22cm}
\footnotesize{$q_\gamma : 221$ MeV $\:\:{\cal S}_{fi} : 9/50 $}&&&&& \\
&&&&& \\
$^3D_2 \rightarrow \chi_{c1}\,\gamma$ & \,\,$0.3920$\,\, & 
\,\,$0.3953$\,\, & \,\,$0.4145$\,\, & 262 keV & {\bf 198 keV} \\
\vspace{-.22cm}
\footnotesize{$q_\gamma : 263$ MeV $\:\:{\cal S}_{fi} : 9/10 $}&&&&& \\
&&&&& \\
$^3D_1 \rightarrow \chi_{c2}\,\gamma$ & \,\,$0.4216$\,\, & 
\,\,$0.4246$\,\, & \,\,$0.4372$\,\, & 5.06 keV & {\bf 4.13 keV} \\
\vspace{-.22cm}
\footnotesize{$q_\gamma : 206$ MeV $\:\:{\cal S}_{fi} : 1/50 $}&&&&& \\
&&&&& \\
$^3D_1 \rightarrow \chi_{c1}\,\gamma$ & \,\,$0.3963$\,\, & 
\,\,$0.3997$\,\, & \,\,$0.4164$\,\, & 123 keV & {\bf 94.9 keV} \\
\vspace{-.22cm}
\footnotesize{$q_\gamma : 248$ MeV $\:\:{\cal S}_{fi} : 1/2 $}&&&&& \\
&&&&& \\
$^3D_1 \rightarrow \chi_{c0}\,\gamma$ & \,\,$0.3578$\,\, & 
\,\,$0.3619$\,\, & \,\,$0.3889$\,\, & 370 keV & {\bf 251 keV} \\
\vspace{-.22cm}
\footnotesize{$q_\gamma : 336$ MeV $\:\:{\cal S}_{fi} : 2 $}&&&&& \\
\end{tabular}}
\end{table}

\begin{table}[h!]
\centering{
\caption{The E1 transitions from the lightest spin-triplet $D$-wave states
in bottomonium~($b\bar b$). The photon momenta $q_\gamma$ and statistical 
factors ${\cal S}_{fi}$ have been calculated as for Table~\ref{E1Dcc}.}
\label{E1Dbb}
\vspace{.4cm}
\begin{tabular}{l||r|r|r||c|c}
\multicolumn{1}{c||}{Transition} &
\multicolumn{3}{c||}{${\cal M}_0$\,\,[fm]} & \multicolumn{2}{c}{Width} \\
&\multicolumn{3}{c||}{} & \multicolumn{2}{c}{} \\
&\multicolumn{1}{c|}{IA} & \multicolumn{1}{c|}{DYN} &
\multicolumn{1}{c||}{E1} & \,E1\, & {\bf DYN} \\
\hline\hline
&&&&& \\
$^3D_3 \rightarrow \chi_{b2}\,\gamma$ & \,\,$-0.1270$\,\, & 
\,\,$-0.1271$\,\, & \,\,$-0.1291$\,\, & 24.5 keV & {\bf 22.3 keV} \\
\vspace{-.22cm}
\footnotesize{$q_\gamma : 242$ MeV $\:\:{\cal S}_{fi} : 18/25 $}&&&&& \\
&&&&& \\
$^3D_2 \rightarrow \chi_{b2}\,\gamma$ & \,\,$-0.1275$\,\, & 
\,\,$-0.1276$\,\, & \,\,$-0.1295$\,\, & 5.51 keV & {\bf 5.04 keV} \\
\vspace{-.22cm}
\footnotesize{$q_\gamma : 233$ MeV $\:\:{\cal S}_{fi} : 9/50 $}&&&&& \\
&&&&& \\
$^3D_2 \rightarrow \chi_{b1}\,\gamma$ & \,\,$-0.1228$\,\, & 
\,\,$-0.1229$\,\, & \,\,$-0.1250$\,\, & 19.6 keV & {\bf 17.8 keV} \\
\vspace{-.22cm}
\footnotesize{$q_\gamma : 253$ MeV $\:\:{\cal S}_{fi} : 9/10 $}&&&&& \\
&&&&& \\
$^3D_1 \rightarrow \chi_{b2}\,\gamma$ & \,\,$-0.1280$\,\, & 
\,\,$-0.1280$\,\, & \,\,$-0.1297$\,\, & 0.54 keV & {\bf 0.50 keV} \\
\vspace{-.22cm}
\footnotesize{$q_\gamma : 223$ MeV $\:\:{\cal S}_{fi} : 1/50 $}&&&&& \\
&&&&& \\
$^3D_1 \rightarrow \chi_{b1}\,\gamma$ & \,\,$-0.1234$\,\, & 
\,\,$-0.1235$\,\, & \,\,$-0.1254$\,\, & 9.75 keV & {\bf 8.89 keV} \\
\vspace{-.22cm}
\footnotesize{$q_\gamma : 243$ MeV $\:\:{\cal S}_{fi} : 1/2 $}&&&&& \\
&&&&& \\
$^3D_1 \rightarrow \chi_{b0}\,\gamma$ & \,\,$-0.1173$\,\, & 
\,\,$-0.1174$\,\, & \,\,$-0.1196$\,\, & 17.3 keV & {\bf 15.5 keV} \\
\vspace{-.22cm}
\footnotesize{$q_\gamma : 275$ MeV $\:\:{\cal S}_{fi} : 2 $}&&&&& \\
\end{tabular}}
\end{table}

\newpage

\begin{table}[h!]
\centering{
\caption{The E1 transitions from the $2P$ states in charmonium~($c\bar 
c$) and bottomonium~($b\bar b$). The labeling of the states and the photon 
momenta $q_\gamma$ for the transitions are as for Tables~\ref{ccE1} 
and~\ref{bbE1}.}
\label{E1rest}
\vspace{.4cm}
\begin{tabular}{l||r|r|r|r||c|c}
\multicolumn{1}{c||}{Transition} &
\multicolumn{3}{c|}{${\cal M}_0$\,\,[fm]} & 
\multicolumn{1}{c||}{${\cal M}_2$\,\,[fm]}
&\multicolumn{2}{c}{Width} \\
&\multicolumn{3}{c|}{}& &\multicolumn{2}{c}{} \\
&\multicolumn{1}{c|}{IA} & \multicolumn{1}{c|}{DYN} &
\multicolumn{1}{c|}{E1} & \,\, & \,E1\, & {\bf DYN} \\
\hline\hline
&&&&&& \\
$\chi_{c2}(2P)\rightarrow \psi\,'\gamma$ & \,\,0.3702\,\, & \,\,0.3735\,\, & 
\,\,0.4007\,\, & $1.09\cdot 10^{-2}$ & 194 keV & {\bf 144 keV} \\
\vspace{-.23cm}
\footnotesize{$q_\gamma : 236$ MeV}&&&&&& \\
&&&&&& \\
$\chi_{c1}(2P)\rightarrow \psi\,'\gamma$ & \,\,0.3996\,\, & \,\,0.4031\,\, & 
\,\,0.4271\,\, & $9.60\cdot 10^{-3}$ & 178 keV & {\bf 137 keV} \\
\vspace{-.23cm}
\footnotesize{$q_\gamma : 220$ MeV}&&&&&& \\
&&&&&& \\
$\chi_{c0}(2P)\rightarrow \psi\,'\gamma$ & \,\,0.4321\,\, & \,\,0.4357\,\, & 
\,\,0.4543\,\, & $7.43\cdot 10^{-3}$ & 133 keV & {\bf 108 keV} \\
\vspace{-.23cm}
\footnotesize{$q_\gamma : 193$ MeV}&&&&&& \\
&&&&&& \\
$\chi_{c2}(2P)\rightarrow J/\psi\,\gamma$ & \,\,0.0660\,\, & \,\,0.0693\,\, & 
\,\,0.0526\,\, & $-6.49\cdot 10^{-3}$ & 133 keV & {\bf 132 keV} \\
\vspace{-.23cm}
\footnotesize{$q_\gamma : 745$ MeV}&&&&&& \\
&&&&&& \\
$\chi_{c1}(2P)\rightarrow J/\psi\,\gamma$ & \,\,0.0549\,\, & \,\,0.0581\,\, & 
\,\,0.0380\,\, & $-7.83\cdot 10^{-3}$ & 65.3 keV & {\bf 90.1 keV} \\
\vspace{-.23cm}
\footnotesize{$q_\gamma : 731$ MeV}&&&&&& \\
&&&&&& \\
$\chi_{c0}(2P)\rightarrow J/\psi\,\gamma$ & \,\,0.0387\,\, & \,\,0.0417\,\, & 
\,\,0.0183\,\, & $-9.13\cdot 10^{-3}$ & 13.6 keV & {\bf 44.9 keV} \\
\vspace{-.23cm}
\footnotesize{$q_\gamma : 707$ MeV}&&&&&& \\
&&&&&& \\
$h_c(2P)\rightarrow \eta_c'\,\gamma$ & \,\,0.3444\,\, & \,\,0.3450\,\,  & 
\,\,0.3752\,\, & $1.21\cdot 10^{-2}$ & 236 keV & {\bf 167 keV}  \\
\vspace{-.23cm}
\footnotesize{$q_\gamma : 263$ MeV $\:\:{\cal S}_{fi} : 3 $}&&&&&& \\
&&&&&& \\
$h_c(2P)\rightarrow \eta_c\,\gamma$ & \,\,$0.0646$\,\, & \,\,$0.0631$\,\, 
& \,\,$0.0489$\,\, & $-5.44\cdot 10^{-3}$ & 161 keV & {\bf 142 keV}   \\
\vspace{-.23cm}
\footnotesize{$q_\gamma : 821$ MeV $\:\:{\cal S}_{fi} : 3 $}&&&&&& \\
&&&&&& \\
\hline
&&&&&& \\
$\chi_{b2}(2P)\rightarrow \Upsilon(2S)\,\gamma$ & \,\,$-0.1186$\,\, & 
\,\,$-0.1186$\,\, & 
\,\,$-0.1221$\,\, & $-1.38\cdot 10^{-3}$ & 18.5 keV & {\bf 16.4 keV} \\
\vspace{-.23cm}
\footnotesize{$q_\gamma : 243$ MeV}&&&&&&
\footnotesize{Br: $16.2\pm 2.4\:\%$} \\
&&&&&& \\
$\chi_{b1}(2P)\rightarrow \Upsilon(2S)\,\gamma$ & \,\,$-0.1240$\,\, & 
\,\,$-0.1241$\,\, & 
\,\,$-0.1272$\,\, & $-1.23\cdot 10^{-3}$ & 16.8 keV & {\bf 15.1 keV} \\
\vspace{-.23cm}
\footnotesize{$q_\gamma : 229$ MeV}&&&&&&
\footnotesize{Br: $21\pm 4\:\%$} \\
&&&&&& \\
$\chi_{b0}(2P)\rightarrow \Upsilon(2S)\,\gamma$ & \,\,$-0.1305$\,\, & 
\,\,$-0.1306$\,\, & 
\,\,$-0.1331$\,\, & $-1.01\cdot 10^{-3}$ & 13.5 keV & {\bf 12.3 keV} \\
\vspace{-.23cm}
\footnotesize{$q_\gamma : 207$ MeV}&&&&&&
\footnotesize{Br: $4.6\pm 2.1\:\%$} \\
&&&&&& \\
$\chi_{b2}(2P)\rightarrow \Upsilon\,\gamma$ & \,\,$-0.0176$\,\, & 
\,\,$-0.0177$\,\, & 
\,\,$-0.0156$\,\, & $8.44\cdot 10^{-4}$ & 10.7 keV & {\bf 11.4 keV} \\
\vspace{-.23cm}
\footnotesize{$q_\gamma : 777$ MeV}&&&&&&
\footnotesize{Br: $7.1\pm 1.0\:\%$} \\
&&&&&& \\
$\chi_{b1}(2P)\rightarrow \Upsilon\,\gamma$ & \,\,$-0.0155$\,\, & 
\,\,$-0.0156$\,\, & 
\,\,$-0.0132$\,\, & $9.28\cdot 10^{-4}$ & 7.32 keV & {\bf 8.40 keV} \\
\vspace{-.23cm}
\footnotesize{$q_\gamma : 764$ MeV}&&&&&&
\footnotesize{Br: $8.5\pm 1.3\:\%$} \\
&&&&&& \\
$\chi_{b0}(2P)\rightarrow \Upsilon\,\gamma$ & \,\,$-0.0123$\,\, & 
\,\,$-0.0124$\,\, & 
\,\,$-0.0098$\,\, & $1.01\cdot 10^{-3}$ & 3.70 keV & {\bf 4.93 keV} \\
\vspace{-.23cm}
\footnotesize{$q_\gamma : 743$ MeV}&&&&&&
\footnotesize{Br: $0.9\pm 0.6\:\%$} \\
&&&&&& \\
$h_b(2P)\rightarrow \eta_b(2S)\,\gamma$ & \,\,$-0.1120$\,\, & 
\,\,$-0.1120$\,\,  & 
\,\,$-0.1155$\,\, & $-1.42\cdot 10^{-3}$ & 19.5 keV & {\bf 17.2 keV}  \\
\vspace{-.23cm}
\footnotesize{$q_\gamma : 257$ MeV $\:\:{\cal S}_{fi} : 3 $}&&&&&& \\
&&&&&& \\
$h_b(2P)\rightarrow \eta_b\,\gamma$ & \,\,$-0.0179$\,\, & 
\,\,$-0.0179$\,\,  & 
\,\,$-0.0160$\,\,  & $7.43\cdot 10^{-3}$ & 13.4 keV & {\bf 13.5 keV}  \\
\vspace{-.23cm}
\footnotesize{$q_\gamma : 821$ MeV $\:\:{\cal S}_{fi} : 3 $}&&&&&& \\
\end{tabular}}
\end{table}

\newpage

\begin{table}[h!]
\centering{
\caption{The E1 dominated transitons between low-lying states in the 
$B_c^\pm$ ($c\bar b$) system. The labeling of the columns is as for 
Tables~\ref{ccE1} and~\ref{bbE1}. Note that spin-triplet states are 
indicated by "stars" in their labels. The $q_\gamma$ values have been 
obtained as for Table~\ref{ccE1}. Further E1 transitions can be found in 
Table~\ref{bcrest}.}
\label{bcE1}
\vspace{.3cm}
\begin{tabular}{l||r|r|r|r||c|c} 
\multicolumn{1}{c||}{Transition} &
\multicolumn{3}{c|}{${\cal M}_0$\,\,[fm]} &
\multicolumn{1}{c||}{${\cal M}_2$\,\,[fm]}
&\multicolumn{2}{c}{Width} \\
&\multicolumn{3}{c|}{}& &\multicolumn{2}{c}{} \\
&\multicolumn{1}{c|}{IA} & \multicolumn{1}{c|}{DYN} &
\multicolumn{1}{c|}{E1} & \,\, & \,E1\, & {\bf DYN} \\
\hline\hline 
&&&&&& \\
$B_{c2}^*\rightarrow B_c^*\,\gamma$ & \,\,0.1361\,\, & \,\,0.1386\,\, &
\,\,0.1453\,\, & $2.43\cdot 10^{-3}$ & 120 keV & {\bf 93.9 keV} \\
\vspace{-.23cm}
\footnotesize{$q_\gamma : 397$ MeV}&&&&&& \\  
&&&&&& \\
$B_{c1}^*\rightarrow B_c^*\,\gamma$ & \,\,0.1382\,\, & \,\,0.1408\,\, &  
\,\,0.1469\,\, & $2.16\cdot 10^{-3}$ & 107 keV & {\bf 84.6 keV} \\
\vspace{-.23cm}
\footnotesize{$q_\gamma : 379$ MeV}&&&&&& \\
&&&&&& \\
$B_{c0}^*\rightarrow B_c^*\,\gamma$ & \,\,0.1403\,\, & \,\,0.1429\,\, &
\,\,0.1480\,\, & $1.81\cdot 10^{-3}$ & 86.6 keV & {\bf 70.4 keV} \\
\vspace{-.23cm}
\footnotesize{$q_\gamma : 352$ MeV}&&&&&& \\
&&&&&& \\
$B_c^*(2S)\rightarrow B_{c0}^*\,\gamma$ & \,\,$-0.1578$\,\, &
\,\,$-0.1578$\,\, &
\,\,$-0.1610$\,\, & $-1.28\cdot 10^{-3}$ & 4.71 keV & {\bf 4.22 keV} \\
\vspace{-.23cm}
\footnotesize{$q_\gamma : 184$ MeV}&&&&&& \\
&&&&&& \\
$B_c^*(2S)\rightarrow B_{c1}^*\,\gamma$ & \,\,$-0.1720$\,\, &
\,\,$-0.1720$\,\, &
\,\,$-0.1746$\,\, & $-1.01\cdot 10^{-3}$ & 10.2 keV & {\bf 9.35 keV} \\
\vspace{-.23cm}
\footnotesize{$q_\gamma : 157$ MeV}&&&&&& \\
&&&&&& \\
$B_c^*(2S)\rightarrow B_{c2}^*\,\gamma$ & \,\,$-0.1842$\,\, &
\,\,$-0.1843$\,\, &
\,\,$-0.1864$\,\, & $-0.85\cdot 10^{-3}$ & 13.2 keV & {\bf 12.3 keV} \\
\vspace{-.23cm}
\footnotesize{$q_\gamma : 139$ MeV}&&&&&& \\
&&&&&& \\
$B_{c1}\rightarrow B_c\,\gamma$ & \,\,0.1267\,\, & \,\,0.1283\,\,  &
\,\,0.1353\,\, & $2.50\cdot 10^{-3}$ & 135 keV & {\bf 103 keV}  \\
\vspace{-.23cm}
\footnotesize{$q_\gamma : 431$ MeV $\:\:{\cal S}_{fi} : 3 $}&&&&&& \\
&&&&&& \\
$B_c(2S)\rightarrow B_{c1}\,\gamma$ & \,\,$-0.1834$\,\, & \,\,$-0.1835$\,\,
& \,\,$-0.1854$\,\, & $-0.73\cdot 10^{-3}$ & 20.7 keV & {\bf 19.3 keV}   \\
\vspace{-.22cm}
\footnotesize{$q_\gamma : 133$ MeV $\:\:{\cal S}_{fi} : 3 $}&&&&&& \\
\end{tabular}}
\end{table}

\begin{table}[h!]
\centering{
\caption{The M1 transitions between low-lying $S$-wave states in the
$B_c^\pm$~($c\bar b$) system. The matrix elements of the spin-flip operator 
associated with the OGE interaction are given in the column labeled "Oge". 
The $3S$ states have not been considered as they lie above the $BD$ 
fragmentation threshold.}
\label{M1tabBc}
\vspace{.3cm}
\begin{tabular}{l||r|r|r|r||c|c|c}
\multicolumn{1}{c||}{Transition} &
\multicolumn{4}{c||}{Matrix element\,\,[$10^{-3}\:$fm]} & 
\multicolumn{3}{c}{Width} \\
&\multicolumn{4}{c||}{}&\multicolumn{3}{c}{} \\
& \multicolumn{1}{c|}{NRIA} & \multicolumn{1}{c|}{RIA} &
\multicolumn{1}{c|}{Conf} & \multicolumn{1}{c||}{Oge}  
&\,NRIA\, & \,RIA\, & {\bf RIA+Exch} \\
\hline\hline
&&&&&&& \\
$B_c^*\rightarrow B_c\,\gamma$ & $18.51\:\,$ & $14.96\:\,$ & $-3.643\:\:$ & 
$3.969$
& 50.0 eV & 32.6 eV & {\bf 34.0 eV} \\
\vspace{-.22cm}
\footnotesize{$q_\gamma : 53$ MeV}&&&&&&& \\
&&&&&&& \\
$B_c^*(2S)\rightarrow B_c\,\gamma$ & $1.015\:\,$ & $-1.437\:\,$ & $1.342\:\:$ & 
$1.182$
& 179 eV & 360 eV & {\bf 206 eV} \\
\vspace{-.22cm}
\footnotesize{$q_\gamma : 576$ MeV}&&&&&&& \\
&&&&&&& \\
$B_c^*(2S)\rightarrow B_c(2S)\,\gamma$ & $18.49\:\,$ & $14.80\:\,$ & $-7.666\:\:$ & 
$2.480$ 
& 3.61 eV & 2.31 eV & {\bf 0.98 eV} \\
\vspace{-.22cm}
\footnotesize{$q_\gamma : 22$ MeV}&&&&&&& \\
&&&&&&& \\
$B_c(2S)\rightarrow B_c^*\,\gamma$ & $-1.067\:\,$ & $-3.089\:\,$ & $1.924\:\:$ & 
$8.351$ 
& 411 eV & 3.44 keV & {\bf 39.5 eV} \\
\vspace{-.22cm}
\footnotesize{$q_\gamma : 507$ MeV}&&&&&&& \\
\end{tabular}}
\end{table}

\newpage

\begin{table}[h!]
\centering{
\caption{The E1 transitions from the $2P$ states in the $B_c^\pm$~($c\bar 
b$) system. The labeling of the states and the photon momenta $q_\gamma$ 
for the transitions are as for Tables~\ref{ccE1} and~\ref{bbE1}.}
\label{bcrest}
\vspace{.3cm}
\begin{tabular}{l||r|r|r|r||c|c}
\multicolumn{1}{c||}{Transition} &
\multicolumn{3}{c|}{${\cal M}_0$\,\,[fm]} & 
\multicolumn{1}{c||}{${\cal M}_2$\,\,[fm]}
&\multicolumn{2}{c}{Width} \\
\vspace{-.13cm}
&\multicolumn{3}{c|}{}& &\multicolumn{2}{c}{} \\
&\multicolumn{1}{c|}{IA} & \multicolumn{1}{c|}{DYN} &
\multicolumn{1}{c|}{E1} & \,\, & \,E1\, & {\bf DYN} \\
\hline\hline
\vspace{-.13cm}
&&&&&& \\
$B_{c2}^*(2P)\rightarrow B_c^*(2S)\,\gamma$ & \,\,0.2151\,\, & 
\,\,0.2166\,\, & 
\,\,0.2265\,\, & $3.94\cdot 10^{-3}$ & 49.3 keV & {\bf 41.7 keV} \\
\vspace{-.26cm}
\footnotesize{$q_\gamma : 222$ MeV}&&&&&& \\
&&&&&& \\
$B_{c1}^*(2P)\rightarrow B_c^*(2S)\,\gamma$ & \,\,0.2263\,\, & 
\,\,0.2279\,\, & 
\,\,0.2370\,\, & $3.59\cdot 10^{-3}$ & 46.5 keV & {\bf 39.9 keV} \\
\vspace{-.26cm}
\footnotesize{$q_\gamma : 212$ MeV}&&&&&& \\
&&&&&& \\
$B_{c0}^*(2P)\rightarrow B_c^*(2S)\,\gamma$ & \,\,0.2381\,\, & 
\,\,0.2398\,\, & 
\,\,0.2474\,\, & $3.04\cdot 10^{-3}$ & 39.0 keV & {\bf 34.2 keV} \\
\vspace{-.26cm}
\footnotesize{$q_\gamma : 194$ MeV}&&&&&& \\
&&&&&& \\
$B_{c2}^*(2P)\rightarrow B_c^*\,\gamma$ & \,\,0.0320\,\, & 
\,\,0.0333\,\, & 
\,\,0.0264\,\, & $-2.97\cdot 10^{-3}$ & 27.0 keV & {\bf 32.9 keV} \\
\vspace{-.26cm}
\footnotesize{$q_\gamma : 733$ MeV}&&&&&& \\
&&&&&& \\
$B_{c1}^*(2P)\rightarrow B_c^*\,\gamma$ & \,\,0.0270\,\, & 
\,\,0.0283\,\, & 
\,\,0.0204\,\, & $-3.34\cdot 10^{-3}$ & 15.5 keV & {\bf 23.2 keV} \\
\vspace{-.26cm}
\footnotesize{$q_\gamma : 723$ MeV}&&&&&& \\
&&&&&& \\
$B_{c0}^*(2P)\rightarrow B_c^*\,\gamma$ & \,\,0.0207\,\, & 
\,\,0.0220\,\, & 
\,\,0.0131\,\, & $-3.69\cdot 10^{-3}$ & 5.95 keV & {\bf 13.4 keV} \\
\vspace{-.26cm}
\footnotesize{$q_\gamma : 707$ MeV}&&&&&& \\
&&&&&& \\
$B_{c1}(2P)\rightarrow B_c(2S)\,\gamma$ & \,\,0.2065\,\, & \,\,0.2076\,\,  & 
\,\,0.2179\,\, & $4.08\cdot 10^{-3}$ & 53.3 keV & {\bf 44.5 keV}  \\
\vspace{-.26cm}
\footnotesize{$q_\gamma : 234$ MeV $\:\:{\cal S}_{fi} : 3 $}&&&&&& \\
&&&&&& \\
$B_{c1}(2P)\rightarrow B_c\,\gamma$ & \,\,$0.0316$\,\, & \,\,$0.0323$\,\,  & 
\,\,$0.0259$\,\, & $-2.75\cdot 10^{-3}$ & 30.6 keV & {\bf 35.7 keV}   \\
\vspace{-.26cm}
\footnotesize{$q_\gamma : 771$ MeV $\:\:{\cal S}_{fi} : 3 $}&&&&&& \\
\end{tabular}}
\end{table}

\begin{table}[h!]
\centering{
\caption{The E1 transitions from the lightest spin-triplet $D$-wave states
in the $B_c^\pm$~($c\bar b$) system. The labeling of the states, the photon 
momenta $q_\gamma$ and statistical factors ${\cal S}_{fi}$ are as for 
Table~\ref{E1Dcc}.}
\label{E1Dbc}
\vspace{.3cm}
\begin{tabular}{l||r|r|r||c|c}
\multicolumn{1}{c||}{Transition} &
\multicolumn{3}{c||}{${\cal M}_0$\,\,[fm]} & \multicolumn{2}{c}{Width} \\
\vspace{-.13cm}
&\multicolumn{3}{c||}{} & \multicolumn{2}{c}{} \\
&\multicolumn{1}{c|}{IA} & \multicolumn{1}{c|}{DYN} &
\multicolumn{1}{c||}{E1} & \,E1\, & {\bf DYN} \\
\hline\hline
\vspace{-.13cm}
&&&&& \\
$^3D_3 \rightarrow B_{c2}^*\,\gamma$ & \,\,$0.2270$\,\, & 
\,\,$0.2285$\,\, & \,\,$0.2352$\,\, & 75.7 keV & {\bf 65.4 keV} \\
\vspace{-.26cm}
\footnotesize{$q_\gamma : 235$ MeV $\:\:{\cal S}_{fi} : 18/25 $}&&&&& \\
&&&&& \\
$^3D_2 \rightarrow B_{c2}^*\,\gamma$ & \,\,$0.2279$\,\, & 
\,\,$0.2294$\,\, & \,\,$0.2358$\,\, & 18.3 keV & {\bf 15.9 keV} \\
\vspace{-.26cm}
\footnotesize{$q_\gamma : 232$ MeV $\:\:{\cal S}_{fi} : 9/50 $}&&&&& \\
&&&&& \\
$^3D_2 \rightarrow B_{c1}^*\,\gamma$ & \,\,$0.2181$\,\, & 
\,\,$0.2197$\,\, & \,\,$0.2267$\,\, & 63.9 keV & {\bf 54.7 keV} \\
\vspace{-.26cm}
\footnotesize{$q_\gamma : 250$ MeV $\:\:{\cal S}_{fi} : 9/10 $}&&&&& \\
&&&&& \\
$^3D_1 \rightarrow B_{c2}^*\,\gamma$ & \,\,$0.2288$\,\, & 
\,\,$0.2303$\,\, & \,\,$0.2362$\,\, & 1.84 keV & {\bf 1.61 keV} \\
\vspace{-.26cm}
\footnotesize{$q_\gamma : 224$ MeV $\:\:{\cal S}_{fi} : 1/50 $}&&&&& \\
&&&&& \\
$^3D_1 \rightarrow B_{c1}^*\,\gamma$ & \,\,$0.2193$\,\, & 
\,\,$0.2209$\,\, & \,\,$0.2274$\,\, & 32.5 keV & {\bf 28.0 keV} \\
\vspace{-.26cm}
\footnotesize{$q_\gamma : 243$ MeV $\:\:{\cal S}_{fi} : 1/2 $}&&&&& \\
&&&&& \\
$^3D_1 \rightarrow B_{c0}^*\,\gamma$ & \,\,$0.2079$\,\, & 
\,\,$0.2096$\,\, & \,\,$0.2170$\,\, & 54.4 keV & {\bf 45.9 keV} \\
\vspace{-.26cm}
\footnotesize{$q_\gamma : 270$ MeV $\:\:{\cal S}_{fi} : 2 $}&&&&& \\
\end{tabular}}
\end{table}

\newpage

\begin{table}[h!]
\parbox{.6\textwidth}{\centering{
\begin{tabular}{l||r|r|r}
& \multicolumn{1}{c|}{$B_c^*$} & \multicolumn{1}{c|}{$B_c^*(2S)$} 
& \multicolumn{1}{c}{$B_c^*(3S)$} \vspace{-.2cm}
\\ &&& \\ 
\hline\hline
&&& \\
NRIA  & 0.4810 $\mu_N$   & 0.4810 $\mu_N$   & 0.4810 $\mu_N$  	  \\ &&& \\
RIA   & 0.4135 $\mu_N$   & 0.4075 $\mu_N$   & 0.3973 $\mu_N$	  \\ &&& \\
Conf  & $-0.0166$ $\mu_N$ & $-0.0342$ $\mu_N$ & $-0.0487$ $\mu_N$ \\ &&& \\
Oge   & 0.0292 $\mu_N$   & 0.0186 $\mu_N$   & 0.0146 $\mu_N$	  \\ &&& \\ 
\hline &&& \\
Total & {\bf 0.426 $\mu_N$} & {\bf 0.392 $\mu_N$} & {\bf 0.363 $\mu_N$}  \\
\end{tabular}}}
\parbox{.38\textwidth}{
\caption{The magnetic moments of the $S$-wave $c\bar b$ states in units of 
nuclear magnetons. The RIA magnetic moment is given by eq.~(\ref{RIAmu}), 
and the exchange current contributions from the scalar confining and OGE
interactions by eqs.~(\ref{Confmu}) and~(\ref{OGEmu}), respectively. Note 
that the \mbox{NRIA} results are equivalent to those given by the static 
quark model.}
\label{Mtab}}
\end{table}

\section{Discussion}

It is instructive to compare the numerical results of the previous 
section both with experiment and with those of other theoretical 
calculations, as there are several issues that are not readily apparent by 
casual inspection of the data presented in Tables~\ref{M1tab} 
through~\ref{Mtab}. A direct comparison with the experimental averages 
given in ref.~\cite{PDG} may be misleading, as the branching fractions for 
various transitions are typically better known than the total width of the 
decaying state. Also, as recently published results~\cite{Hagiwara} for 
$\chi_{cJ}\rightarrow J/\psi\,\gamma$ indicate that these E1 widths are 
significantly larger than previously thought~\cite{PDG}, then a review of 
the model predictions is called for.

\subsection{M1 transitions in the $c\bar c$ system}

\begin{itemize}
\item $J/\psi\rightarrow \eta_c\,\gamma$

The observed width of $1.14\,\pm\,0.39$ keV for the M1 transition 
$J/\psi\rightarrow \eta_c\,\gamma$ has been difficult to explain 
theoretically~\cite{Shifmanpap}, since previous calculations typically 
overestimate it by a factor~$\sim 3$. A possible solution for this 
overprediction, already hinted at in ref.~\cite{GrOld}, is presented in 
Table~\ref{M1tab}, where the exchange current contribution from the scalar 
confining interaction brings the width down to the desired level. However, 
as shown in ref.~\cite{Oldlahde}, expansion of the RIA spin-flip operator 
to order $v^2/c^2$ overestimates the correction to the static quark model 
result, and in that case the usefulness of the exchange 
current contribution is not apparent. The importance of negative energy 
components for the transition $J/\psi\rightarrow \eta_c\,\gamma$ has also 
been established within the instantaneous approximation to the 
Bethe-Salpeter equation in ref.~\cite{Snellman} and the Schr\"odinger 
approach in ref.~\cite{Oldlahde}, where widths close to that given in 
Table~\ref{M1tab} were obtained for a scalar confining interaction. If the 
whole $Q\bar Q$ potential had effective vector coupling structure, then 
no exchange current contribution would arise, as a vector interaction 
contributes a spin-flip operator only if the quark and antiquark masses are 
unequal, and agreement with experiment would thus be excluded. Other 
possible solutions include the introduction of a large anomalous magnetic 
moment for the charm quark~\cite{GrOld}, but this possibility has 
apparently not been substantiated.

\item $\psi\,'\rightarrow \eta_c\,\gamma$

This nonrelativistically forbidden M1 transition has also proved 
theoretically challenging, since the (near) orthogonality of the quarkonium 
wavefunctions leads to model-dependent results. In the recent calculation 
by ref.~\cite{Snellman}, where good agreement with experiment was found for 
$J/\psi\rightarrow \eta_c\,\gamma$, the width for $\psi\,'\rightarrow 
\eta_c\,\gamma$ was however overpredicted by almost an order of magnitude. 
The present calculation gives a width of $\sim 1.1$ keV for that 
transition, which is close to the empirical value $0.84\,\pm\,0.24$ 
keV~\cite{Hagiwara}. This result is, however, not robust. That such a 
favorable result is obtained depends delicately on several factors, such as 
the employment of $\psi\,'$ and $\eta_c$ wavefunctions that include the 
spin-spin interaction in the $S$-wave, and the choice of approximation for 
the M1 matrix element. 

The amplitude~(\ref{matrM1}) has the 
advantage of allowing the use of a realistic photon momentum in the 
expression~(\ref{M1dec}) for the M1 width. It is useful to note that this 
treatment yields the same spin-flip operators than had the rigorous M1 
approximation been used, as in the calculation of the exchange magnetic 
moment operators in refs.~\cite{Helm-PhD,Tsushima}. Furthermore, the M1 
approximation has been taken to affect the entire factor in brackets in 
eq.~(\ref{matrM1}). If the exponentials were separated from the current 
operators in eq.~(\ref{matrM1}), then the width for $\psi\,'\rightarrow 
\eta_c\,\gamma$ would be overpredicted by a factor $\sim 4$. However, if 
spin-averaged wavefunctions were employed, as in ref.~\cite{Oldlahde}, then 
the conclusion would be exactly the opposite; In that case the present 
treatment would lead to unfavorable results. As seen from 
Table~\ref{M1tab}, the exchange current operator associated with the scalar 
confining interaction gives the main contribution to the width for 
$\psi\,'\rightarrow \eta_c\,\gamma$ within this calculation. The present 
treatment of the M1 approximation may be regarded as consistent since it 
leads to the correct spin-flip operators and simultaneously allows the 
recoil of the $\eta_c$ to be taken into account. Further modifications to 
the M1 width for $\psi\,'\rightarrow \eta_c\,\gamma$ may result from a 
relativistic treatment of the exchange current operator~(\ref{2qcurr}) and 
consideration of possible $D$-wave admixture in the $\psi'$ state due to 
the OGE tensor interaction.

\end{itemize}

\subsection{M1 transitions in the $b\bar b$ and $c\bar b$ systems}

\begin{itemize}
\item $\Upsilon\rightarrow \eta_b\,\gamma$

A reasonable prediction for the width of the M1 transition 
$\Upsilon\rightarrow \eta_b\,\gamma$ can be obtained as soon as the mass of 
the $\eta_b$ state is constrained by experiment, as the contributions from 
two-quark operators are suppressed by the large mass of the $b$ quark. The 
unceratinty in the constituent mass~(4885 MeV) of the $b$ quark is also 
nowadays rather small~\cite{Quigg,GrE1bb}. As realistic models of the 
spin-spin splittings for $S$-wave quarkonia give an $\eta_b$ mass around 
9400 MeV, then the width for $\Upsilon\rightarrow \eta_b\,\gamma$ is likely 
to be less than 10 eV, as given in Table~\ref{M1tab}. However, the 
reliability of such predictions cannot be tested until the $\eta_b$ state 
is discovered empirically.

\item $B_c^*\rightarrow B_c\,\gamma$

At this time, the photon momenta involved in calculations of M1 widths in 
the $c\bar b$ system have to be extracted from model calculations of the 
$c\bar b$ mass spectrum. As shown in ref.~\cite{Quigg}, the mass of the 
spin triplet $B_c^*$ state is rather well constrained, and is expected to 
be about 6350 MeV. However, there is in general no such agreement for the 
magnitude of the $B_c^* - B_c$ splitting, which determines the photon 
momentum of the $B_c^*\rightarrow B_c\,\gamma$ transition. Realistic models 
for the spin-spin interaction in the $S$-wave appear to favor a small 
splitting of $\sim 40$ MeV. In spite of this, the computed width for 
$B_c^*\rightarrow B_c\,\gamma$ of 34 eV given in Table~\ref{M1tabBc} 
compares well with the $\sim 29$ eV predicted by ref.~\cite{Quigg}, even 
though exchange current contributions were not considered in that work. It 
is noteworthy that the exchange current contribution from the OGE 
interaction is dominant for $c\bar b$ and leads to a slight increase of 
the width for $B_c^*\rightarrow B_c\,\gamma$. Because of this cancellation, 
the nonrelativistic width is quite close to the net result. 

\end{itemize}
\pagebreak

Predictions have also been given, in Tables~\ref{M1tab} and~\ref{M1tabBc}, 
for M1 transitions between $Q\bar Q$ states that lie below the respective 
fragmentation thresholds. Notable among these is $\psi\,'\rightarrow 
\eta_c'\,\gamma$, which is similar to $J/\psi\,\rightarrow \eta_c \,\gamma$. 
However, the suppression due to the scalar confining interaction is 
stronger as the $2S$ wavefunctions have a longer range. Also, as recent 
experimental results indicate that the mass of the $\eta_c'$ is larger than 
previously thought~\cite{Belle}, then the phase space available for 
$\psi\,'\rightarrow \eta_c'\,\gamma$ is also suppressed. Consequently the 
predicted width is also smaller than the values suggested by 
previous work~\cite{Snellman}. As for the transition $\psi\,'\rightarrow 
\eta_c\,\gamma$, the width for 
$\eta_c'\rightarrow J/\psi\,\gamma$ is also sensitive to the particulars of 
the model because of cancellations in the matrix element and the large 
photon momentum. The results in Table~\ref{M1tab} suggest that the width 
for this transition should be around 2 keV. As the experimental situation 
concerning the $\eta_c'$ continues to improve, then the width for 
$\eta_c'\rightarrow J/\psi\,\gamma$ may possibly be measured in the near 
future. 

In the case of the $b\bar b$ system, the number of measurable M1 
transitions is 
larger since the $3S$ states of bottomonium lie below the threshold 
for $B\bar B$ fragmentation. The results of Table~\ref{M1tab} indicate that 
most M1 transitions in the $b\bar b$ system are difficult to calculate 
accurately, and therefore provide an instructive test for model 
predictions. In particular, the widths for transitions that do not change 
the principal quantum number of the quarkonium state are predicted to be 
highly suppressed, whereas the widths for transitions from excited $\eta_b$ 
states to the $\Upsilon$ ground state are predicted to have larger widths 
of about 100~eV.

\subsection{E1 transitions in the $c\bar c$ system}

While the M1 transitions in heavy quarkonia are sensitive to the 
Lorentz structure of the $Q\bar Q$ interaction, the E1 transitions have 
been shown here to receive only small contributions from the exchange 
current operators of Fig.~\ref{feyn}. They typically increase the value of 
the matrix element so that its value in the dynamical model is between those in 
the impulse and E1 approximations. On the other hand, the 
matrix element of the dipole operator is sensitive to the shape of the 
$Q\bar Q$ wavefunctions, and thus a realistic description of the E1 widths 
requires that the hyperfine components of the $Q\bar Q$ interaction are not 
treated as first order perturbations. This conclusion is in line with that 
reached in ref.~\cite{GrE1bb}, which employed the nonsingular $Q\bar Q$ 
potential model of ref.~\cite{Gupta}. Comparisons between the present 
results, those of previous calculations and experiment are given 
for several E1 transitions in $c\bar c$ and $b\bar b$ in 
Tables~\ref{restabc} and~\ref{restabb}.

\begin{table}[h!]
\centering{
\caption{Comparison of the results for E1 transitions in the 
charmonium ($c\bar c$) system with the predictions of other models, for a 
scalar confining interaction. All widths are given in keV. The 
experimental widths have been extracted from ref.~\cite{Hagiwara}.}
\label{restabc}
\vspace{.4cm}
\begin{tabular}{c||c|c|c|c|r}
& MB (ref.~\cite{McClary}) & GS (ref.~\cite{GrOld,Grhc}) 
& MR (ref.~\cite{Rosner})  & {\bf This Work} & Exp (ref.~\cite{Hagiwara}) \\ 
\vspace{-.3cm}
&&&&& \\
\hline\hline 
\vspace{-.2cm}
&&&&& \\
$\chi_{c2}\rightarrow J/\psi\,\gamma$ & 347 & 413 & 609 & {\bf 343} 
& $389 \pm 60$ keV \\
$\chi_{c1}\rightarrow J/\psi\,\gamma$ & 270 & 340 & 460 & {\bf 276} 
& $290 \pm 60$ keV \\ 
$\chi_{c0}\rightarrow J/\psi\,\gamma$ & 128 & 162 & 225 & {\bf 144}  
& $165 \pm 40$ keV \\ 
\vspace{-.2cm}
&&&&& \\
\hline
\vspace{-.2cm}
&&&&& \\
$\psi\,'\rightarrow \chi_{c2}\,\gamma$ & 27 & 26 & 41 & {\bf 33} 
& $20.4 \pm 4.0$ keV \\
$\psi\,'\rightarrow \chi_{c1}\,\gamma$ & 31 & 28 & 48 & {\bf 39} 
& $25.2 \pm 4.5$ keV \\
$\psi\,'\rightarrow \chi_{c0}\,\gamma$ & 19 & 18 & 37 & {\bf 33} 
& $26.1 \pm 4.5$ keV \\ 
\vspace{-.2cm}
&&&&& \\
\hline
\vspace{-.2cm}
&&&&& \\
$h_c \rightarrow \eta_c\,\gamma$      & 483 & 630 & --  & {\bf 370} 
& -- \hspace{.9cm} \\ 
\end{tabular}}
\end{table}

\newpage

\begin{itemize}
\item{$\chi_{cJ}\rightarrow J/\psi\,\gamma$}

Predictions for the E1 transitions from the spin-triplet $P$-wave states in 
charmonium should be robust, as the quarkonium wavefunctions involved do 
not contain any nodes. The systematic overprediction of these widths by 
$\sim 20\%$, as indicated by the empirical data of ref.~\cite{PDG}, has 
therefore been puzzling. However, in view of the most recent experimental 
data~\cite{Hagiwara}, this issue is apparently resolved. As seen from 
Table~\ref{restabc}, models which consider the recoil of the $J/\psi$ 
reproduce the empirical results remarkably well. In contrast, the rigorous 
E1 approximation leads to overpredictions of~\mbox{$\sim 50\%$.} 

\item{$\psi\,'\rightarrow \chi_{cJ}\,\gamma$}

The E1 transitions from the $\psi'$ state have, in general, been difficult 
to predict, since the E1 approximation has typically overestimated the 
empirical widths by at least a factor $\sim 2$. The latest experimental 
data for the branching fractions for $\gamma$ decay and total width of the 
$\psi'$ suggests that the widths for $\psi\,'\rightarrow \chi_{cJ}\,\gamma$ 
should be around 25~keV, while the E1 approximation typically yields widths 
in excess of 40 keV. The results for $\psi\,'\rightarrow \chi_{cJ}\,\gamma$ 
obtained by ref.~\cite{Snellman} within the framework of the instantaneous 
approximation to the Bethe-Salpeter equation, are also of the order $\sim 
40$ keV. Recoil effects cannot account for this overprediction, 
as they are much weaker than for the $\chi_{cJ}\rightarrow J/\psi\,\gamma$ 
transitions. However, inspection of Table~\ref{restabc} reveals that the 
predicted relative widths are also not in agreement with experiment, 
although the uncertainties are considerable. This suggests a sensitivity to 
small changes in the $Q\bar Q$ wavefunctions, which is indeed revealed by 
inspection of the matrix elements in Table~\ref{ccE1}. The present 
empirical data suggests that the width for $\psi\,'\rightarrow 
\chi_{c2}\,\gamma$ should be the smallest. Of the models presented in 
Table~\ref{restabc}, the present one comes closest to this result. It 
should also be noted that significant reductions of the E1 widths have 
been obtained in ref.~\cite{Eichtenqq} by consideration of closed $c\bar q 
- q\bar c$ fragmentation channels. 

\end{itemize}

The predicted widths for E1 transitions from the $\psi(3S)$   
state have also been given in Table~\ref{ccE1}. The results suggest that 
the widths for transitions to the $2P$ states should be comparable to 
those for the $\psi\,'\rightarrow \chi_{cJ}\,\gamma$ transitions. On the 
other hand, the widths for the $\psi(3S)\rightarrow \chi_{cJ}\,\gamma$ 
transitions are predicted to be smaller by a factor~$\sim 4$. The empirical 
detection of any of these transitions will probably be challenging since 
the $\psi(3S)$ state decays mainly through $D\bar D$ fragmentation. 
However, the photon produced in the $h_c\rightarrow \eta_c\gamma$ 
transition will probably be detected in the near future, as that state lies 
well below the $D\bar D$ fragmentation threshold. The results presented in 
Table~\ref{restabc} suggest that the E1 width for that transition is the 
largest in the $c\bar c$ system. However, the uncertainty in that 
prediction is somewhat larger because of recoil effects and the hitherto 
uncertain mass of the $h_c$ state. Of particular interest are also the E1 
transitions from the $^3D_1$ state given in Table~\ref{E1Dcc}, as that 
state probably corresponds to the empirical $\psi(3770)$ resonance. The 
calculated widths suggest that the transitions to the $\chi_{c1}$ and 
$\chi_{c0}$ states should be detectable by experiment, whereas that to the 
$\chi_{c2}$ state is highly suppressed by the statistical factor 
${\cal S}_{fi}$.

\subsection{E1 transitions in the $b\bar b$ system}

The number of measurable E1 transitions in the $b\bar b$ system is larger 
than in the $c\bar c$ system, as the $\Upsilon(3S)$ state lies below the 
threshold for $B\bar B$ fragmentation. While the branching fractions 
for several of the transitions presented in Table~\ref{restabb} have been 
measured, only very few of the total widths are known. This is due to the 
narrowness of the $b\bar b$ states, which makes a direct determination of 
their widths difficult. It is therefore instructive to compare the 
predicted E1 widths with those of other models, as well as with experiment. 
A review of the most important E1 transitions in Table~\ref{restabb} is 
given below.

\newpage

\begin{table}[h!]
\centering{
\caption{Comparison of the predictions for E1 transitions in the $b\bar b$ 
system with those of other models that use a scalar confining interaction. 
All widths are given in keV. The experimental widths have been extracted 
from ref.~\cite{PDG}. Note that only the branching fractions are known for 
most of the transitions.}
\label{restabb}
\vspace{.4cm}
\begin{tabular}{c||c|c|c|c}
& GS (ref.~\cite{GrOld}) & GZ (ref.~\cite{GrE1bb}) 
& {\bf This Work} & Exp (ref.~\cite{PDG}) \\ 
\vspace{-.3cm}
&&&& \\
\hline\hline 
\vspace{-.2cm}
&&&& \\
$\chi_{b2}\rightarrow \Upsilon\,\gamma$ & 33.0 & 33.8 & {\bf 36.0} & 
$22 \pm 3$\% \\
$\chi_{b1}\rightarrow \Upsilon\,\gamma$ & 29.8 & 30.4 & {\bf 32.5} & 
$35 \pm 8$\% \\
$\chi_{b0}\rightarrow \Upsilon\,\gamma$ & 25.7 & 25.3 & {\bf 26.6} & 
$< 6$ \% \\
\vspace{-.2cm}
&&&& \\
\hline
\vspace{-.2cm}
&&&& \\
$\Upsilon'\rightarrow \chi_{b0}\,\gamma$ & 0.73 & 0.76 & {\bf 1.01} & 
$1.7 \pm 0.5$ keV \\
$\Upsilon'\rightarrow \chi_{b1}\,\gamma$ & 1.62 & 1.37 & {\bf 1.80} & 
$3.0 \pm 0.7$ keV \\
$\Upsilon'\rightarrow \chi_{b2}\,\gamma$ & 1.84 & 1.45 & {\bf 2.03} & 
$3.1 \pm 0.7$ keV \\
\vspace{-.2cm}
&&&& \\
\hline
\vspace{-.2cm}
&&&& \\
$\chi_{b2}'\rightarrow \Upsilon'\,\gamma$ & 12.9 & 16.2 & {\bf 16.4} 
& $16.4 \pm 2.4$ \% \\
$\chi_{b1}'\rightarrow \Upsilon'\,\gamma$ & 11.9 & 14.7 & {\bf 15.1} 
& $21 \pm 4$ \% \\
$\chi_{b0}'\rightarrow \Upsilon'\,\gamma$ & 10.6 & 12.3 & {\bf 12.3} 
& $4.6 \pm 2.1$ \% \\
\vspace{-.2cm}
&&&& \\
\hline
\vspace{-.2cm}
&&&& \\
$\chi_{b2}'\rightarrow \Upsilon\,\gamma$ & 18.2 & 10.4 & {\bf 11.4} 
& $7.1 \pm 1.0$ \% \\
$\chi_{b1}'\rightarrow \Upsilon\,\gamma$ & 11.8 & 7.51 & {\bf 8.40} 
& $8.5 \pm 1.3$ \% \\
$\chi_{b0}'\rightarrow \Upsilon\,\gamma$ & 6.50 & 3.57 & {\bf 4.93} 
& $0.9 \pm 0.6$ \% \\
\vspace{-.2cm}
&&&& \\
\hline
\vspace{-.2cm}
&&&& \\
$\Upsilon''\rightarrow \chi_{b0}\,\gamma$ & 0.114 & 0.029 & {\bf 0.15} 
& -- \\
$\Upsilon''\rightarrow \chi_{b1}\,\gamma$ & 0.003 & 0.095 & {\bf 0.11} 
& -- \\
$\Upsilon''\rightarrow \chi_{b2}\,\gamma$ & 0.194 & 0.248 & {\bf 0.04} 
& -- \\
\vspace{-.2cm}
&&&& \\
\hline
\vspace{-.2cm}
&&&& \\
$\Upsilon''\rightarrow \chi_{b0}'\,\gamma$ & 1.09 & 1.30 & {\bf 1.14} 
& $1.4 \pm 0.3$ keV \\
$\Upsilon''\rightarrow \chi_{b1}'\,\gamma$ & 2.15 & 2.34 & {\bf 2.12} 
& $3.0 \pm 0.5$ keV \\
$\Upsilon''\rightarrow \chi_{b2}'\,\gamma$ & 2.29 & 2.71 & {\bf 2.50}  
& $3.0 \pm 0.6$ keV \\
\end{tabular}}
\end{table}

\begin{itemize}
\item $\chi_{bJ}\rightarrow \Upsilon\,\gamma$

The calculated widths for $\chi_{bJ}\rightarrow \Upsilon\,\gamma$ agree 
rather well with those of the previous calculations presented in 
Table~\ref{restabb}, although they are generally somewhat larger. The 
results of the present calculation suggest that the total width of the 
$\chi_{b2}$ state should be $164\pm 22$ keV and that of the $\chi_{b1}$ 
$93\pm 22$ keV. Similarly, the total width of the $\chi_{b0}$ is probably 
larger than $\sim 440$ keV. This situation is similar to that observed for 
$c\bar c$~\cite{PDG}, where the $\chi_{c2}$ is wider than the $\chi_{c1}$ 
by about a factor $\sim 2$.

\item $\Upsilon'\rightarrow \chi_{bJ}\,\gamma$

The largest uncertainty concerning the $\Upsilon(2S)\rightarrow 
\chi_{bJ}\,\gamma$ is the measurement of the total width of the 
$\Upsilon(2S)$ state. Originally given as $\sim 27$ keV, it now stands at 
$44\pm 7$ keV~\cite{Hagiwara}. Thus the model predictions in 
Table~\ref{restabb}, which originally fitted the experimental data well, no 
longer do so satisfactorily. It is therefore very difficult to judge the 
quality of any given prediction until the experimental situation is 
stabilized. Still, it is noteworthy that the present calculation does give 
slightly better agreement with experiment than the previous models in 
Table~\ref{restabb}.

\newpage

\item {\bf Transitions from the $\chi_{bJ}'$ states}

The E1 transitions from the $\chi_{bJ}(2P)$ states in bottomonium provide a 
useful test for theoretical models since experimental data now exists on 
all six branching fractions~\cite{PDG}, even though the total widths of the  
$\chi_{bJ}(2P)$ states are not known. The experimental results indicate 
that the widths for transitions to the $\Upsilon$ should be about one half 
of those for transitions to the $\Upsilon(2S)$, even though much more phase 
space is available for the former. Indeed, it can be seen from 
Table~\ref{restabb} that the model of ref.~\cite{GrOld}, where 
spin-averaged wavefunctions were employed, does not compare very well with 
the experimental branching fractions even though the hyperfine splittings 
of the $\chi_{bJ}(2P)$ states are quite small. A realistic description of 
these transitions requires that the hyperfine effects are accounted for by 
the $Q\bar Q$ wavefunctions, in which case the computed E1 widths agree much 
better with experiment. 

As the calculated widths for $\chi_{bJ}(2P)\rightarrow 
\Upsilon(2S)\,\gamma$ are almost the same in ref.~\cite{GrE1bb} and the 
present model, then it is possible to obtain realistic estimates for the 
total widths of the $\chi_{bJ}(2P)$ states from the measured branching 
fractions for the E1 transitions $\chi_{bJ}(2P)\rightarrow 
\Upsilon(2S)\,\gamma$. The width of the $\chi_{b2}(2P)$ state is then 
predicted to be $100 \pm 15$ keV, while that of the $\chi_{b1}(2P)$ is $72 
\pm 14$ keV. The $\chi_{b0}(2P)$ state appears to be significantly broader, 
but because of the large errors in the reported E1 branching fractions, 
only a rough estimate of $267 \pm 140$ keV is possible.

\item {\bf Transitions from the $\Upsilon''$ state}

As the reported total width of the $\Upsilon(3S)$ state~\cite{PDG}, $26.3 
\pm 3.5$ keV, is better known than that of the $\Upsilon(2S)$ state, then 
it is expected that systematic uncertainties in the reported experimental 
results for $\Upsilon(3S)\rightarrow \chi_{bJ}(2P)\,\gamma$ should be 
smaller than for the analogous $\Upsilon(2S)\rightarrow \chi_{bJ}\,\gamma$ 
transitions. By inspection of Table~\ref{restabb}, it can be seen that the 
$\Upsilon(3S)\rightarrow \chi_{bJ}(2P)\,\gamma$ transitions are generally 
rather well described by a number of models, although the calculation of 
ref.~\cite{GrOld}, where spin-averaged wavefunctions were employed, appears 
to underpredict the empirical widths. Also, the results of 
ref.~\cite{GrE1bb} appear to compare slightly more favorably with 
experiment than does the present calculation. 

On the other hand, the situation concerning the $\Upsilon(3S)\rightarrow 
\chi_{bJ}\,\gamma$ transitions remains unsettled because of a strong 
cancellation in the E1 matrix element. As all of the models presented in 
Table~\ref{restabb} predict different widths for the 
$\Upsilon(3S)\rightarrow \chi_{bJ}\,\gamma$ transitions, then only 
experimental determination of the widths can settle the question. However, 
this may turn out to be a formidable task since all models predict 
widths that are an order of magnitude smaller than that of any previously 
measured E1 transition in the $b\bar b$ system. It is seen by inspection of 
Table~\ref{bbE1} that the calculated widths in the E1 approximation are 
similar to those of ref.~\cite{GrOld} in that the width for 
$\Upsilon(3S)\rightarrow \chi_{b1}\,\gamma$ is vanishingly small. However, 
the dynamical model predicts that the width for $\Upsilon(3S)\rightarrow 
\chi_{b0}\,\gamma$ should be the largest and that for 
$\Upsilon(3S)\rightarrow \chi_{b2}\,\gamma$ the smallest. It is encouraging 
that the same pattern is also predicted in Table~\ref{ccE1} for the 
analogous transitions in the $c\bar c$ system, where the widths are much 
larger relative to the other E1 transitions. It is also noteworthy that the 
$\Upsilon(3S)\rightarrow \chi_{bJ}\,\gamma$ transitions obtain an 
appreciable contribution from the matrix element ${\cal M}_2$ in 
Table~\ref{bbE1}.

\end{itemize}

In addition to the transitions considered in Table~\ref{restabb}, the E1 
decays of the spin-singlet $h_b$ and $\eta_b$ states and the lightest 
spin-triplet $D$-wave states have also been calculated in 
Tables~\ref{bbE1},~\ref{E1Dbb} and~\ref{E1rest}. No empirical data exists 
as yet on any of these states. The pattern of E1 widths for these 
states is predicted to be similar to that for the analogous states in 
the $c\bar c$ system, although the widths are much smaller for $b\bar b$ 
because of the narrower wavefunctions involved.

\newpage

\subsection{E1 transitions in the $c\bar b$ system}

\begin{table}[h!]
\begin{minipage}{0.43\textwidth}
The determination of photon momenta for E1 transitions in the 
bottom-charm $B_c^\pm$ mesons has to rely completely on 
model \mbox{predictions} 
for the masses of the $c\bar b$ states. However, the uncertainty 
introduced by this is \mbox{small}, as model predictions for the major 
level \mbox{splittings} agree with each other to a large 
\mbox{extent~\cite{Quigg}.} The 
results in Table~\ref{restabbc} reveal that the predictions of the 
present model are similar to those obtained by ref.~\cite{Quigg}, although 
significant differences exist for transitions such as $B_c^*(2S)\rightarrow 
B_{c0}^*\,\gamma$ and $B_{c2}^*(2P)\rightarrow B_c^*(2S)\,\gamma$, where 
the widths are sensitive to the effects of the hyperfine components of the 
$Q\bar Q$ interaction. It is noteworthy that while 
the predicted widths for the $B_{cJ}^*\rightarrow B_c^*\,\gamma$ 
transitions agree rather well with those from ref.~\cite{Quigg}, there is a 
significant disagreement for $B_{c1}\rightarrow B_c\,\gamma$. When the 
somewhat different photon momenta are accounted for, this disagreement 
amounts to about a factor~$\sim 3$. An issue not considered in this paper 
is the mixing of the $L=1$ states with $J=1$, which is due to the 
antisymmetric spin-orbit interaction that was not included in the 
Hamiltonian~(\ref{ham}). This mixing, which was considered in 
ref.~\cite{Quigg}, has the effect of allowing "spin-flip" E1 transitions of 
the type $B_{c1}^* \rightarrow B_c\,\gamma$. However, the widths for such 
"forbidden" transitions were found in ref.~\cite{Quigg} to be typically 
suppressed by a factor $\sim 100$ relative to the "allowed" ones considered 
in this work.
\end{minipage}
\hfill
\begin{minipage}{0.53\textwidth} 
\centering{
\caption{Comparison of the results for E1 \mbox{transitions} in the
$B_c^\pm$ ($c\bar b$, $b\bar c$) system with the predictions of 
\mbox{ref.~\cite{Quigg}.} All widths are given in keV. In the notation 
$B_{cJ}^*$, where $J$ is the total angular momentum, states without 
\mbox{stars} in their labels are spin singlets in the $LS$ coupling 
scheme.}
\label{restabbc}
\vspace{.4cm}
\begin{tabular}{c||c|c}
& EQ (ref.~\cite{Quigg}) & {\bf This Work} \\
\vspace{-.3cm}
&& \\
\hline\hline
\vspace{-.2cm}
&& \\
$B_{c2}^*\rightarrow B_c^*\,\gamma$ & 112.6  & {\bf 93.9 } \\
$B_{c1}^*\rightarrow B_c^*\,\gamma$ & 99.5   & {\bf 84.6 } \\
$B_{c0}^*\rightarrow B_c^*\,\gamma$ & 79.2   & {\bf 70.4 } \\
\vspace{-.2cm}
&& \\
\hline
\vspace{-.2cm}
&& \\
${B_c^*}'\rightarrow B_{c0}^*\,\gamma$ & 7.8      & {\bf 4.22 } \\
${B_c^*}'\rightarrow B_{c1}^*\,\gamma$ & 14.5     & {\bf 9.35 } \\
${B_c^*}'\rightarrow B_{c2}^*\,\gamma$ & 17.7     & {\bf 12.3 } \\
\vspace{-.2cm}
&& \\
\hline
\vspace{-.2cm}
&& \\
$B_{c1}\rightarrow B_c\,\gamma$ 	& 56.4 	& {\bf 103  } \\
$B_c\hspace{.007cm}'\rightarrow B_{c1}\,\gamma$ & 5.2	& {\bf 19.3 } \\
\vspace{-.2cm}
&& \\
\hline
\vspace{-.2cm}
&& \\
${B_{c2}^*}'\rightarrow {B_c^*}'\,\gamma$ & 73.8 & {\bf 41.7}  \\
${B_{c1}^*}'\rightarrow {B_c^*}'\,\gamma$ & 54.3 & {\bf 39.9}  \\
${B_{c0}^*}'\rightarrow {B_c^*}'\,\gamma$ & 41.2 & {\bf 34.2}  \\
\vspace{-.2cm}
&& \\
\hline
\vspace{-.2cm}
&& \\
${B_{c2}^*}'\rightarrow B_c^*\,\gamma$ & 25.8	& {\bf 32.9} \\
${B_{c1}^*}'\rightarrow B_c^*\,\gamma$ & 22.1	& {\bf 23.2} \\
${B_{c0}^*}'\rightarrow B_c^*\,\gamma$ & 21.9	& {\bf 13.4} \\
\vspace{-.2cm}
&& \\
\hline
\vspace{-.2cm}
&& \\
$^3D_3 \rightarrow B_{c2}^*\,\gamma$ & 98.7	& {\bf 65.4}	\\
$^3D_2 \rightarrow B_{c2}^*\,\gamma$ & 24.7	& {\bf 15.9}	\\
$^3D_2 \rightarrow B_{c1}^*\,\gamma$ & 88.8	& {\bf 54.7}	\\
$^3D_1 \rightarrow B_{c2}^*\,\gamma$ & 2.7	& {\bf 1.61}	\\
$^3D_1 \rightarrow B_{c1}^*\,\gamma$ & 49.3	& {\bf 28.0}	\\
$^3D_1 \rightarrow B_{c0}^*\,\gamma$ & 88.6	& {\bf 45.9}	\\
\end{tabular}}
\end{minipage}
\end{table}

As the magnetic moment operators of a $Q\bar Q$ system have been derived in 
this work, then the magnetic moment of the $S$-wave spin-triplet $B_c^*$ 
states have been calculated as an interesting by-product. It is seen from 
Table~\ref{Mtab} that the magnetic moments also receive significant 
corrections from the two-quark operators considered in this paper. This is 
not surprising since the situation is similar for the M1 decays of the 
$Q\bar Q$ systems. Inspection of Table~\ref{Mtab} reveals that the net 
relativistic decrease of the $B_c^*$ magnetic moment amounts to about 15 
\%, and that the exchange current contributions actually increase the 
net magnetic moment slightly. This situation was also noted for the M1 
transition $B_c^*\rightarrow B_c\,\gamma$. However, because of the 
short-range nature of the OGE interaction, its contribution quickly 
becomes subdominant for the higher $S$-wave states. The predicted magnetic 
moments of the $B_c^*$ mesons are in line with the work of 
ref.~\cite{Dannbom} on the magnetic moments of the baryons.

\newpage

\subsection{Other frameworks}

The masses, radiative transitions and other properties of the heavy 
quarkonia have also been investigated within the frameworks of QCD sum 
rules~\cite{Colangelo,Kiselev}, Heavy Quark Effective Theory 
(HQET)~\cite{Casa} and Non-Relativistic QCD (NRQCD)~\cite{Bramb}. It is of 
particular interest to compare the results of the present calculation with 
those of ref.~\cite{Casa}, where the E1 transitions between $c\bar c$ and 
$b\bar b$ states were considered. The results of ref.~\cite{Casa} are 
presented in terms of an "effective" matrix element $\delta$, from which 
the dependence on the angular momenta of the $Q\bar Q$ states and the 
momentum of the emitted photon has been extracted. The theoretical 
uncertainties are therefore concentrated into the parameter $\delta$. As 
that parameter is closely related to the radial matrix element ${\cal M}_0$ 
for E1 decay, a direct comparison between the present work and that of 
ref.~\cite{Casa} is possible. Such a comparison is given in 
Table~\ref{deltatab}. In addition, the width for the E1 transitions 
$h_c\rightarrow \eta_c\,\gamma$ was in ref.~\cite{Casa} obtained as $\sim 
450$~keV, which in view of the somewhat larger photon momentum of 
ref.~\cite{Casa} compares reasonably well with the present value of 
370~keV.

\begin{table}[h!]
\parbox{.4\textwidth}{
\caption{Comparison between the effective matrix element $\delta$ of 
ref.~\cite{Casa} and the corresponding values extracted from the present 
calculation. The exact definition of $\delta$ can be found in 
ref.~\cite{Casa}. Note that in the present model, the matrix element for an 
E1 transition also depends on the total angular momentum $J$, as 
spin-averaged wavefunctions are not employed. The reported values in this 
table therefore correspond to the spin average of the matrix elements given 
in Tables~\ref{ccE1} and~\ref{bbE1}.}
\label{deltatab}}
\parbox{.59\textwidth}{
\centering{
\begin{tabular}{c||c|c}
 & ref.~\cite{Casa} [$\mathrm{GeV}^{-1}$] & {\bf This Work} \\ 
&& \vspace{-.2cm} \\ \hline\hline && \\
$\psi'\rightarrow \chi_{cJ}\,\gamma$  & $0.209\pm 0.012$ & {\bf 0.267} \\
&& \\
$\chi_{cJ}\rightarrow J/\psi\,\gamma$ & $0.197\pm 0.012$ & {\bf 0.215} \\
&& \\
$\Upsilon'\rightarrow \chi_{bJ}\,\gamma$ & $0.109\pm 0.009$ & {\bf 0.086} \\
&& \\
$\Upsilon''\rightarrow \chi_{bJ}'\,\gamma$ & $0.152\pm 0.009$ & {\bf 0.138} \\
\end{tabular}}}
\end{table}

\subsection{Conclusions}

The conclusion of this work concerning the exchange current operators and 
the associated two-quark E1 and M1 operators is that they play a major role 
in spin-flip M1 transitions, whereas they are insignificant for the E1 
transitions. The reason for the smallness of the exchange charge 
contributions to the E1 widths is the large masses of the charm and bottom 
quarks, as those contributions are proportional to $m^{-3}$. However, as 
suggested in ref.~\cite{Pidec}, such operators may be significant when 
light ($\sim 400$ MeV) constituent quarks are involved.

A satisfactory description of the M1 widths of charmonium was achieved in 
this work, under the assumption that the effective confining interaction is 
purely scalar. This is in line with the results of ref.~\cite{Snellman}, 
where the width for $J/\psi\rightarrow \eta_c\,\gamma$ was found to be well 
described by an effective scalar confining interaction within the framework 
of the instantaneous approximation to the Bethe-Salpeter equation. This is 
reassuring since it has been shown in ref.~\cite{Gromes} that 
models (of equal mass quarkonia) which employ an effective linear vector 
confining interaction or a superposition of scalar and vector confining 
interactions with positive weights are inconsistent with the properties of 
QCD. 

It should be noted that the 
instanton induced interaction for $Q\bar q$ and $Q\bar Q$ systems, as given 
by ref.~\cite{Zahed}, may also contribute a significant two-quark exchange 
charge operator. For light constituent quarks, such contributions may be 
large whereas they are much smaller for charm quarks~\cite{LahdeDs}. A 
calculation of the associated spin-flip operators for M1 transitions could 
therefore provide useful and constraining information on the strength of 
the effective instanton induced interaction in mesons with heavy quarks.

\vspace{1.5cm}
\centerline{\bf Acknowledgments}
\vspace{0.5cm}
The author expresses his thanks to Prof. Dan-Olof Riska for his valuable 
input on the E1 transitions, to Dr. Christina Helminen for instructive 
conversations concerning the derivation of the exchange charge operators, 
and to the Waldemar von Frenckell foundation for a fund grant during the 
completion of this work.

\end{document}